\documentclass[aps,reprint,floatfix,prx,nofootinbib]{revtex4-2}
\usepackage{graphicx} 

\usepackage{amsmath,amssymb,amsfonts,bm}
\usepackage{accents}
\usepackage{sansmath}
\usepackage{mathtools}

\usepackage{upgreek}
\usepackage{booktabs}

\usepackage{tikz}

\usepackage{xcolor}
\definecolor{linkColor}{RGB}{0,70,120}
\definecolor{darkgreen}{RGB}{0,128,0}
\definecolor{darkgray}{RGB}{90,90,90}
\definecolor{orange}{RGB}{180,50,0}

\usepackage[colorlinks=true, allcolors=linkColor,pdfborder={0 0 0},pdfencoding = auto]{hyperref}

\newcommand{\nc}[1]{{\color{purple}[NC: #1]}}

\usepackage[normalem]{ulem}

\newcommand{\rvec}{\mathbf{r}}

\newcommand{\xivec}{{\bm{\xi}}}
\newcommand{\Df}{{\mathbf{D}\kern-0.08em f}}

\let\Re\relax
\DeclareMathOperator{\Re}{Re}
\let\Im\relax
\DeclareMathOperator{\Im}{Im}

\newcommand{\dev}{\operatorname{dev}}

\renewcommand{\arraystretch}{1.2}

\begin{document}

\title{Emergent Elasticity and Quasiconformal Morphing in Active Solids}

\author{Nikolas H.\ Claussen}
\email{nc1333@princeton.edu}
\affiliation{Princeton Center for Theoretical Science, Princeton University, Princeton, New Jersey 08542, USA}
\author{Fridtjof Brauns}
\email{fbrauns@pks.mpg.de}
\affiliation{Max Planck Institute for the Physics of Complex Systems, Nöthnitzer Straße 38, 01187 Dresden, Germany}
\affiliation{Max Planck Institute of Molecular Cell Biology and Genetics, Pfotenhauerstraße 108, 01307 Dresden, Germany}
\affiliation{Center for Systems Biology Dresden, 01307 Dresden, Germany}
\author{Boris I.\ Shraiman}
\email{shraiman@ucsb.edu}
\affiliation{Kavli Institute for Theoretical Physics, University of California, Santa Barbara, California 93106, USA}

\begin{abstract}
    Elasticity theory rests on a constitutive relation between stress and strain relative to a reference state. Living matter, however, generates forces by motor molecule activity independent of strain, requiring a different approach to active tissue mechanics. We develop a theory of two-dimensional continuum mechanics in which the active force configuration, not a reference shape, is the fundamental input. Motivated by the cell-level Active Tension Network model for epithelia, we encode motor-driven intercellular forces in a Riemannian \emph{tension metric}, whose mapping into physical space relates macroscopic stress to the microscopic tensions. Assuming force balance, we show that a stress-free reference state and an effective stress-strain relation emerge at the tissue scale, making the system an effectively elastic \emph{active solid}. Extending from statics to dynamics, we show how such a solid can morph its shape through adiabatic evolution of active tensions. To capture plastic cell rearrangement, we introduce a second metric representing the cell adjacency graph in the continuum; topological rearrangement then appears as a continuous reparametrization of the tension manifold. The key fields characterizing the tissue's local mechanical state -- tension anisotropy, macroscopic stress, and cell shape -- arise from the discrepancies between the tension, adjacency, and physical metrics, represented mathematically by quasiconformal maps. Our theory explains how local contractility controls tissue shape and yields predictions distinct from existing models: tissues can remain rigid while ``flowing'', with a speed set by the rate of change of active forces rather than their magnitude, and shear flow can occur even under isotropic stress, since macroscopic stress is distinct from local tension anisotropy.
\end{abstract}

\maketitle

\section*{Introduction}

Living tissue is a paradigmatic example of Active Matter. During morphogenesis, cellular activity within tissues sculpts the physical form of the developing embryo.
In contrast to inert materials, living tissues generate forces internally~\cite{Gilbert.Barresi2016, Guillot.Lecuit2013}.
Key challenges are to characterize the macroscopic mechanical properties (i.e.\ the ``phase of matter'') of developing tissues and to understand the control of tissue shape (morphogenesis) by the cell-generated forces.

Conventional elasticity, based on a reference shape and a constitutive law relating stress and strain (rate)~\cite{Landau.Lifshitz1986}, has served as the starting point for much work on tissue mechanics~\cite{Oster.etal1983,Alt.etal2017,Coen.Cosgrove2023}. For instance, epithelial cells are often described as having a target area and perimeter~\cite{Farhadifar.etal2007}, with deformation from a target shape defining the stress.
However, biological observations challenge this paradigm. First, the turnover of the force-bearing molecules in cells, rapid compared to morphogenetic timescales, allows the ``rest length'' of a cell to change~\cite{Agarwal.Zaidel-Bar2019}, ruling out a fixed reference state. Since stress needs to be actively maintained against relaxation due to turnover, one expects active stresses to dominate on long timescales.
Active stress, however, is not controlled by strain but by the activity of molecular motors, such as myosin.
Nevertheless, tissues can maintain a macroscopic shape against external shear forces, and, more remarkably, change shape by internally controlled cell rearrangement during morphogenesis.
We refer to such a system, which has neither a reference state nor a constitutive law, as an \emph{active solid}.

We therefore seek to understand tissue mechanics by starting from the specification of cell-associated active stress instead of a reference shape. The timescale of mechanical equilibration (seconds~\cite{Bambardekar.etal2015}) is well separated from that of morphogenetic remodeling (hours).
One thus expects cells to adjust their shapes and positions to achieve force balance by rearranging local sources of active stress. Morphogenesis takes the form of an adiabatic transformation between force-balanced states, controlled by changes in the local mechanical activity.

More precisely, we are motivated by the study of confluent, two-dimensional epithelia, one of the main types of animal tissue~\cite{Gilbert.Barresi2016}.
During development, epithelia often undergo extensive in-plane shear deformation (e.g., convergent-extension ~\cite{Brauns.etal2024}), as well as bending and folding out-of-plane ~\cite{Martin2020}.
Here, we focus on the active mechanics and plasticity of (nearly) planar tissues. This case is relevant to understanding morphogenesis in, e.g., the early
\emph{Drosophila} and avian embryos~\cite{Irvine.Wieschaus1994,Yang.etal2002}, where the embryonic epithelia are constrained by a rigid eggshell, and has attracted significant interest~\cite{Farhadifar.etal2007,Bi.etal2015,Saw.etal2017,Saadaoui.etal2020}.

In many epithelia, dominant forces are tensions generated by molecular motors associated with the cortical cytoskeleton localized at cell-cell interfaces and linked by apical adherens junctions. These active contractile elements at interfaces act as microscopic muscles~\cite{Noll.etal2017,Kim.etal2021} that form a tissue-wide mechanical network.
This picture is made precise by the active tension network (ATN) model~\cite{Noll.etal2017}, a minimal model for epithelial mechanics.
Within this framework, we can ask how cells control tissue shape through local active tensions.
To build intuition, consider a conventional fluid foam. Here, a ``cell'' (foam bubble) interface has no rest length since its tension is independent of its length, set instead by the fluid's surface tension. These surface tensions are balanced against internal pressure in the foam bubbles. Remarkably, even though microscopically fluid, the foam macroscopically behaves like an elastic solid -- it can maintain its shape against gravity~\cite{Weaire.etal2005}.
In a foam, interfacial tensions are fixed material parameters. In a tissue, they are promoted to active, dynamical variables (under local cellular control), enabling richer behavior.

\subsubsection{Summary of main results}

Here, we propose a continuum theory for the mechanics of a 2D tissue dominated by active tensions. The key technical idea is to represent the internal tension configuration by a Riemannian \emph{tension metric} $\mathbf{g}$, a continuum version of Maxwell--Cremona force tilings~\cite{Baker.McRobie2025, Cremona1879}.
The cell tiling defines a (generally non-isometric) embedding of the tension metric into physical space, and we obtain a \emph{stress-metric relation} for the Cauchy stress generated by the (infinitesimal) active force units.
We analyze this geometric model using complex analysis and quasiconformal maps.
We show that the central elements of elasticity theory, a stress-free reference state and a stress-strain relationship, emerge from this embedding. 
Adiabatic dynamics of the active force configuration change the reference state, leading to plastic flow.
Additionally, we provide a unifying geometric formalism to describe cell-scale topological rearrangements, which enable large deformations, in the continuum theory.
This description naturally captures cell-, tension-, and stress-anisotropies and the links between them.
Taken together, these ideas enable a continuum description of active plastic flow.
In a companion paper~\cite{Claussen.etal2026a}, we show that the continuum-based, ``top-down'' approach presented here precisely matches a ``bottom-up'' coarse-graining analysis of the ATN model. 

Our main finding is that internal, active forces in a tissue define a physical reference shape that elastically resists deformations. This makes the system an \emph{emergent active solid}: it behaves like a liquid at the microscale, but elasticity emerges at the macroscale. We then show how adiabatic dynamics of local active forces allow tissues to ``flow'' while retaining rigidity, i.e., without becoming a fluid macroscopically.
The proposed picture of biological shape control during morphogenesis leads to concrete, experimentally testable predictions for epithelia:
\begin{itemize}
    \item Tissues can retain rigidity as they ``flow''.
    \item The speed of tissue flow is set by the rate at which active
forces change, independently of their overall strength.
    \item The macroscopic active stress is distinct from the local tension anisotropy; tissues can undergo directed flow while the active stress remains isotropic.
    \item Effective macroscopic parameters (e.g., elastic moduli, yield threshold) can be predicted from the underlying microscopics (e.g., tension at interfaces).
\end{itemize}

Our description of epithelial tissue behavior is distinct from the interpretation of tissue flow as a solid-to-fluid (rigidity-loss) transition~\cite{Bi.etal2015,Park.etal2015,Tetley.etal2019,Wang.etal2020}. During this transition, cells are predicted to become ``floppy''\footnote{The interfacial tensions vanish in the fluid phase of the so-called ``area-perimeter'' model due to ``excess'' target perimeter~\cite{Bi.etal2015}.}. This conflicts with experimental observations where non-vanishing cellular tensions drive tissue deformation~\cite{Rauzi.etal2008}.
Many existing continuum theories similarly model tissues as viscous fluids driven by an imbalanced active stress~\cite{Streichan.etal2018,Saadaoui.etal2020,Serra.etal2023}, and the flow speed is set by the timescale over which active forces relax viscously. 
In our theory, tissue flow is instead due to the \textit{dynamics} of \textit{balanced} active forces (tensions). As noted above, this has several experimentally observable consequences which distinguish our theory from active fluid models.
For example, tissues can flow under constant, vanishing macroscopic stress, testable by tissue-scale laser ablation.

More generally, our theory describes a class of ``active solids'' not found in conventional continuum mechanics, and lays the foundation for future work on other active systems, such as mesenchymal tissues.

The remainder of this paper is structured as follows. 
We begin by briefly reviewing the cell-level ATN model, one of the motivations of our continuum theory. By means of a simplified example, we explain how the ATN model can be seen as a ``finite-difference'' discretization of the continuum theory.
In Results Section~\ref{sec:continuum_description}, we describe cell tiling and tension configuration in the continuum limit. Section~\ref{sec:statics} studies the static problem, demonstrating the emergence of an effective stress-free reference state and elastic response to deformations.
Section~\ref{sec:dynamics} turns to the adiabatic dynamics of active stress and shows how local tension dynamics, controlled biologically, translate into large-scale shape change. Finally, Sec.~\ref{sec:topology_anisotropy} considers topological cell rearrangement.
Table~\ref{tab:notation} summarizes the notation.

\begin{figure}
    \centering
    \includegraphics{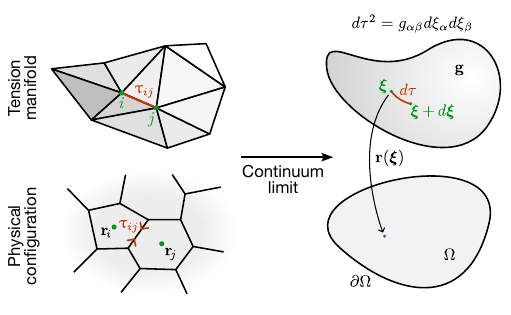}
    \caption{A triangulation (left) is a discrete Riemannian surface.
    The tension metric defines junctional tensions. A map of the Riemannian surface into physical space embeds the tension metric and is the continuum equivalent of the centroids in the cell tiling.
    }
    \label{fig:tri-to-mfd}
\end{figure}

\subsubsection{Discrete theory: 2d tension networks}
\label{sec:ATN-setup}

The microscopic ATN model~\cite{Noll.etal2017} describes a 2d tiling of cells $i,j,\dots$ whose mechanics are dominated by tensions $\tau_{ij}$ along cell-cell interfaces and intracellular pressure $p_i$ (Fig.~\ref{fig:tri-to-mfd}, bottom). Accordingly, the differential of elastic energy is given by 
\begin{align}
    \label{eq:ATN_discrete}
    d E = \sum_{ij} \tau_{ij} d\ell_{ij} - \sum_i p_i dA_i
\end{align}
where $\ell_{ij}, A_i$ are interface lengths and cell areas. In mechanical balance, $dE=0$.
We assume that the cells' positions $\rvec_i$ are constrained to a fixed two-dimensional surface, and that the cells do not exert traction forces on a substrate (therefore, the tissue-internal forces balance among themselves).
This picture applies, for example, to the early \emph{Drosophila} and chicken embryonic epithelia, which sit atop a liquid yolk and are constrained in the third dimension by the eggshell~\cite{Gilbert.Barresi2016}. 
The tensions $\tau_{ij}$ are dynamical variables determined by intrinsic motor molecule activity, independently of the interface length. This is the key premise of the ATN model \cite{Noll.etal2017,Claussen.etal2024}.

By contrast, the pressures $p_i$ are determined by a local equation of state (EoS) $p_i = P(A_i)$ as a function of cell area (the original ATN model \cite{Noll.etal2017} assumed uniform $p_i$).
This is necessary: actively prescribing both tensions and pressures is mechanically unstable.
For example, consider a soap bubble of radius $R$ with an elastic energy $E(R) = 4\pi\tau R^2- \tfrac{4\pi}{3} p R^3$. If the pressure drop $p $ is independent of bubble volume, the bubble is unstable: the Laplace radius $R = 2\tau / p$ is an energy maximum.
Note that a constitutive law for pressure does not imply solid, elastic behavior: fluids also have a pressure and an EoS.

Tissue dynamics unfold on three distinct timescales. The relaxation to mechanical equilibrium is the most rapid timescale, on the order of seconds~\cite{Bambardekar.etal2015}. 
Turnover of force-bearing proteins within a few minutes sets the relaxation speed of passive elastic stress in the cytoskeleton~\cite{McGrath.etal1998}. Finally, morphogenetic remodeling and regulation of contractile activity take place over tens of minutes to hours.  
Here, we exploit this hierarchy and consider only the dynamics at the slowest, morphogenetic timescale. We therefore assume that the ATN model is in instantaneous mechanical equilibrium and that transient passive stresses relax rapidly.
In this limit, the interfacial tensions $\tau_{ij}$ are purely determined by local motor activity and \emph{independent} of interface lengths $\ell_{ij}$ -- there are no reference lengths and no constitutive laws for tensions.

In this stress-first setting, we aim to understand how active tensions determine the physical configuration of the cell tiling in force balance, and how this configuration evolves when tensions change adiabatically. To this end, it will be useful to think of the tension configuration as a network dual to the cell tiling. Since the cells generically meet at 3-fold vertices, the tensions form an intrinsically defined triangulation with one node per cell $i$ and a length $\tau_{ij}$ edge between neighbors $i$ and $j$; Fig.~\ref{fig:tri-to-mfd}.
Each tricellular vertex $ijk$ corresponds to one ``tension triangle'', the shape of which encodes the local tension configuration (for example, tension anisotropy). This Maxwell-Cremona-like ``tension triangulation'' serves as a point of departure for our analysis.

\subsubsection{Introductory example: the rectangular foam}

\begin{figure}[h]
    \centering
    \includegraphics[width=0.55\linewidth]{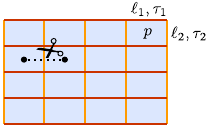}
    \caption{``Rectangular foam'' with anisotropic surface tensions $\tau_1$ on horizontal edges and $\tau_2$ on vertical edges. The stress tensor measures forces across ``virtual cuts'' (dashed line). 
    }
    \label{fig:rect-foam}
\end{figure}

We next discuss a toy model that connects the discrete ATN model to the continuum theory we will present in this manuscript. It generalizes the ``square foam'' introduced in Ref.~\cite{Alexander1998} and is composed of identical rectangular cells.
This minimal example already captures key elements of actual tension networks and foams (which are locally hexagonal and non-uniform), as analyzed in the companion paper~\cite{Claussen.etal2026a}. 
The rectangular cells have an internal pressure $p$ and possibly anisotropic interfacial tensions $\tau_1, \tau_2$ (Fig.~\ref{fig:rect-foam}).
Force balance requires $dE = \tau_{1}d\ell_1 +\tau_2d\ell_2 -pd(\ell_1\ell_2) = 0$. For a pressure $p_0$, the equilibrium lengths are $\ell^{\kern0.1em0}_{1,2} = \tau_{2,1}/p_0$.
Thus, the tensions control the tissue's aspect ratio, and the pressure controls the overall size.

The macroscopic mechanical state is described by the total Cauchy stress tensor $\bm{\sigma}^\mathrm{tot}$. It measures the traction force across a (virtual) cut with length $\ell$ and normal $\hat{\mathbf{n}}$, $\mathbf{f}/\ell = \bm{\sigma}^\mathrm{tot} \cdot \hat{\mathbf{n}}$ (Fig.~\ref{fig:rect-foam}).
Stress arises from the tensions $\tau_1,\tau_2$ (for cuts perpendicular to the respective edges) and the pressure $p$ (contributing isotropically):
\begin{align}
    \label{eq:Batchelor_rectangular_main}
    \bm{\sigma}^\mathrm{tot} = \begin{pmatrix} \frac{\tau_1}{\ell_2} -p & 0 \\ 0 & \frac{\tau_2}{\ell_1} -p \end{pmatrix} = p_0 \begin{pmatrix} \frac{\ell^{\kern0.05em0}_{2}}{\ell_{2}}- \frac{p}{p_0} & 0 \\ 0 & \frac{\ell^{\kern0.05em0}_{1}}{\ell_{1}} - \frac{p}{p_0}  \end{pmatrix}
\end{align}
Thus, for $\ell_{1,2}^{} = \ell^{\kern0.1em0}_{1,2}$ and $p = p_0$, the system is stress-free, $\bm{\sigma}^\mathrm{tot}=0$. (The purpose of distinguishing a particular \emph{reference pressure} $p_0$ from the current pressure $p$ will become clear in the next paragraph; effectively, $p_0$ will act as a dimensional constant for the overall scale of mechanical forces.)
In this unloaded state, the tensile stress is isotropic (even though the tensions $\tau_1, \tau_2$ are anisotropic), and canceled by pressure.
Thus, an effective stress-free reference state emerges from the tension configuration. 
Externally imposed deformations change the lengths and lead to a non-zero total stress, the hallmark of elastic behavior (even though the tensions $\tau_1, \tau_2$ are length-independent).

Recognizing the (diagonal) strain tensor in Eq.~\eqref{eq:Batchelor_rectangular_main}, one reads off the \emph{effective} shear modulus as $p_0/2$. (Off-diagonal shear cannot be described by this minimal model.)
Remarkably, the elastic response is isotropic, despite the microscopic anisotropy. 
The bulk modulus receives two contributions: from the change of ``tension density'' $\tau/\ell$, and from the change of pressure $p$, linked by the EoS $p = P(n)$ to cell density $n = 1/(\ell_1 \ell_2)$. We linearize the EoS around the cell density $n_0 = 1/(\ell^{\kern0.1em0}_{1}\ell^{\kern0.1em0}_{2})$ of the unloaded state, 
$P(n) = p_0 + p_0 B(\tfrac{n}{n_0}-1)$. The cell-level (dimensionless) bulk modulus $B$ then contributes to the effective bulk modulus, which reads $-p_0/2 + p_0 B$. Stability requires $B > 1/2$. Note that for positive tensions $\tau_{1,2} > 0$, the effective shear modulus $p_0/2$ is always positive, set by the reference pressure, and independent of the functional form of $P(n)$.

To connect with the continuum theory, we introduce material coordinates $\bm{\xi}$ enumerating rows and columns of cells. In these coordinates, we can encode the edge tensions in a tension metric
\begin{equation}
\renewcommand{\arraystretch}{1.5}
    \begin{array}{@{} c @{\quad} c @{}}
  \begin{array}[c]{c}
    \tau_{1}^2 = g_{\alpha\beta}  \; d\xi_{2}^\alpha d\xi_{2}^\beta\\
    \tau_{2}^2 = g_{\alpha\beta}  \; d\xi_{1}^\alpha d\xi_{1}^\beta
  \end{array}
  &
  \Rightarrow \quad \mathbf{g}(\bm{\xi}) = \begin{pmatrix} \tau_2^2 & 0 \\ 0 & \tau_1^2\end{pmatrix}.
\end{array}
\end{equation}
Note that $\tau_{2}$ appears in the $(1,1)$ entry since the tension metric encodes tension on edges \emph{perpendicular} to the displacement vector $d\bm{\xi}$.
In the physical configuration, specified by $\ell_1,\ell_2$, the metric reads
\begin{align}
    \label{eq:g_rectangular_main}
    \mathbf{g}(\rvec) = \begin{pmatrix} \tau_2^2/\ell_{1}^2 & 0 \\ 0 & \tau_1^2/\ell_{2}^2\end{pmatrix}.
\end{align}
Eqs.~\eqref{eq:Batchelor_rectangular_main} and~\eqref{eq:g_rectangular_main} lead to a stress-metric relation:
\begin{align}
    \bm{\sigma}^\mathrm{tot} =  \bm{\epsilon}^T \! \cdot \mathbf{g}(\rvec)^{1/2} \cdot \bm{\epsilon} - p\mathbb{I}
\end{align}
where $\bm{\epsilon}$ is the antisymmetric (Levi--Civita) tensor. Note that in the unloaded reference configuration $\ell_{1,2}^0$, the tension metric, and hence the tensile stress, is isotropic. 
We shall refer to such isotropic configurations as \emph{isothermal}\footnote{
This term derives from differential geometry and has no connection to physical temperature~\cite{Lee2018}.}.

\section*{Results}
\setcounter{subsection}{0}

\subsection{Continuum description of ATNs by Riemannian geometry}{\label{sec:continuum_description}}

\subsubsection{Tension and induced metrics}
\label{sec:continuum_derivation}

In the continuum theory, a set of Lagrangian material coordinates $\bm{\xi}$ plays the role of the cell labels $i,j, \dots$.
A \emph{physical configuration} $\rvec(\xivec)$ maps material elements to physical positions (akin to cell centroids in the discrete theory; Fig.~\ref{fig:tri-to-mfd}, bottom).
For simplicity, we assume that physical space is the flat plane, but the theory readily generalizes to tissues on arbitrary curved (but static) surfaces (App.~\ref{app:curved}; e.g., the ellipsoidal surface defined by the \emph{Drosophila} eggshell).
The physical configuration defines the physical distance between material elements, which, in Lagrangian coordinates, can be measured by the \emph{induced metric}
\begin{align}
    h_{\alpha\beta} = \frac{\partial r_{\gamma}}{\partial \xi_\alpha}\frac{\partial r_{\gamma}}{\partial \xi_\beta}
\end{align}

Next, we need to describe the internal configuration of active forces in the tissue. To do so, we draw on the representation of the tensions as a triangulation.
A triangulation, intrinsically specified by its adjacency graph and edge lengths, defines a (piecewise-linear) metric surface\footnote{
This construction is well-defined as long as the microscopic tensions form non-degenerate triangles.
The definition of the manifold structure and the continuum limit are explained in detail in our companion paper~\cite{Claussen.etal2026a}.
If tensions at a vertex violate the triangle inequality, this construction is not possible; however, as we will see below, such configurations are rapidly resolved by topological cell rearrangement.} (see Fig.~\ref{fig:tri-to-mfd}, top).
We can therefore coarse-grain the tension triangulation to a continuous \emph{tension metric} $g_{\alpha\beta}(\bm{\xi})$. 
For an infinitesimal line element $d\bm{\xi}$, the tension metric yields the tension $d\tau$ on cell interfaces oriented \emph{transverse}
to it as
\begin{align}
    d\tau^2 = g_{\alpha\beta}(\bm{\xi}) d\xi_\alpha d\xi_\beta.
\end{align}
In the ``microscopic limit''
where $d\bm{\xi}$ connects two adjacent triangulation nodes, $d\tau$ is the tension along the interface between the corresponding cells.
The tension metric and the induced metric have complementary roles: the former defines the active force and the latter the physical distance between adjacent material elements.

One can think of the physical configuration $\rvec(\xivec)$ as an embedding of the tension metric, that is, in general, not isometric (length-preserving).
Specifically, under the map $\rvec(\xivec)$, the tension metric $\mathbf{g}$ transforms as
\begin{equation} \label{eq:metric_transform}
    g_{\alpha\beta}(\rvec) = \frac{\partial \xi_\gamma}{\partial r_\alpha}g_{\gamma\delta}(\bm{\xi}) \frac{\partial \xi_\delta}{\partial r_\beta}
\end{equation}
Motivated by the introductory ``rectangular foam'' example, we anticipate that there will be a configuration $\rvec(\xivec)$ where tensional stress becomes isotropic and is canceled by an isotropic pressure (uniform, $p=p_0$, for a flat tension
metric).
It is, therefore, convenient to non-dimensionalize $\mathbf{g} \mapsto p_0^{-2} \mathbf{g}$, where the dimensional factor $p_0$ converts units of line tension to units of length.

Different triangulations (i.e., local tension configurations) can realize the same tension manifold. For instance, all triangulations of the plane realize the Euclidean metric. 
The tension triangulation thus contains ``microscopic'' information not encoded in the tension metric $\mathbf{g}$; we return to this point systematically in Sec.~\ref{sec:topology_anisotropy}.
One important additional degree of freedom is the local cell density $n(\xivec)$, i.e., the number of cells (triangulation nodes) per unit area.
The cell density determines the continuum pressure field $p$ (the equivalent of the intracellular pressures $p_i$) via an EoS $p=P(n)$. In contrast to the active tensions, we hence treat pressure as a \emph{passive} field. As we argued in the Introduction, this is necessary for mechanical stability.
Under $\rvec(\xivec)$, the density transforms as $n(\rvec) = n(\xivec) / {\det[\partial_\xivec \rvec]}$, and therefore:
\begin{align}\label{eq:pressure_constitutive}
     p(\rvec)=P \bigl( n(\xivec) / \det[\partial_\xivec \rvec] \bigr).
\end{align}

\subsubsection{The stress-metric relation}

\begin{figure}
    \centering
    \includegraphics{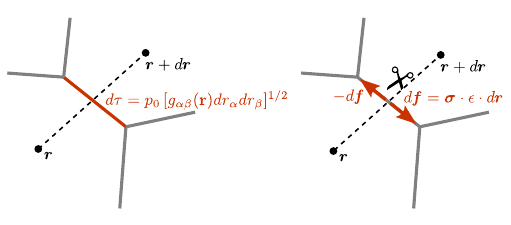}
    \caption{
    Left: In the continuum limit, the displacement vector between two cell centroids becomes a differential $d\mathbf{r}$.
    The tension metric determines the tension $d\tau$ on the interface between these cells. In isothermal coordinates, $dr= d\tau / (p_0\lambda_g)$. 
    Right: The stress tensor $\bm{\sigma}$ is defined by the traction forces that act through infinitesimal virtual cuts. A cut along $d\rvec$ must yield the interfacial tension, relating tension metric and stress tensor in Eq.~\eqref{eq:stress-metric-relation}.}
    \label{fig:virtual_cut}
\end{figure}

We now link the geometry of the tension metric and its mapping to physical space to mechanics. The macroscopic Cauchy stress tensor of an active solid is the sum of two contributions
\begin{align}
    \sigma^\mathrm{tot}_{\alpha\beta} = \sigma_{\alpha\beta} - p \, \delta_{\alpha\beta}
    \label{eq:sigma_tot}
\end{align}
where $p \, \delta_{\alpha\beta}$ is the isotropic pressure, and $\sigma_{\alpha\beta}$ is the tensile stress due to the $\tau_{ij}$.

The Cauchy stress tensor relates the infinitesimal force $d\mathbf{f} = \bm{\sigma} \cdot \hat{\mathbf{n}} \, dr$ acting on a virtual cut with the cut's unit normal $\hat{\mathbf{n}}$ and length $dr$ in the physical configuration.
To see how the tension metric determines $\bm{\sigma}$, consider two adjacent cells at $\rvec$ and $\rvec+d\rvec$ (we implicitly set $dr$ to be of the scale of a single cell, infinitesimal in the continuum limit).
The magnitude of the force $d\bm{f}$ acting through the virtual cut $d\rvec$ (Fig.~\ref{fig:virtual_cut}) equals the tension $|d\bm{f}|=d\tau$ along the interface between the two cells. It is determined by the tension metric, $d\tau^2 = p_0^2 \, g_{\alpha\beta} (\rvec) dr_\alpha dr_\beta$.
On the other hand, the stress tensor applied to the cut normal yields $d\bm{f} = \bm{\sigma}\cdot \hat{\mathbf{n}}\, dr  = \bm{\sigma}\cdot \bm{\epsilon} \cdot d\rvec$, where $\bm{\epsilon}$ is the antisymmetric tensor and $\cdot$ is the standard Euclidean dot product. Equating tension- and cell-space expressions:
\begin{align}
     p_0^2 \, d\rvec^T \cdot \mathbf{g}(\rvec) \cdot d\rvec &= d\tau^2 = d\rvec^T \cdot (\bm{\epsilon}^T \! \cdot \bm{\sigma}^2(\rvec) \cdot \bm{\epsilon} ) \cdot d \rvec \nonumber \\
     \Rightarrow \; \bm{\sigma}^2(\rvec) &= p_0^2 \,  \bm{\epsilon}^T \! \cdot \mathbf{g}(\rvec) \cdot \bm{\epsilon} 
\end{align}
The second line follows because equality holds for arbitrary $d\rvec$.
The stress is tensile, and thus must physically be positive definite.
Therefore, the stress tensor is the (positive) matrix square root of the metric, rotated by $\pi/2$, and scaled by $p_0$:
\begin{align}
    \label{eq:stress-metric-relation}
    \bm{\sigma}(\rvec) =  p_0 \bm{\epsilon}^T \! \cdot \mathbf{g}(\rvec)^{1/2} \cdot \bm{\epsilon} 
\end{align}
This is exactly the relation anticipated by the rectangular foam model in the Introduction. As we will see below, it implies a positive shear modulus, guaranteed by the positive definiteness of $\mathbf{g}$.
We refer to  Eq.~\eqref{eq:stress-metric-relation} as the \emph{stress-metric relation}. It links the microscopic stress, encoded in the tension metric, to the macroscopic stress tensor, defined in physical space. Thus, it replaces the stress-strain relation of a conventional elastic solid.
The square root can be rationalized purely on grounds of dimensional analysis, since the tension metric has dimension $[\mathrm{tension}]^2/[\mathrm{length}]^{2}$ and the stress tensor 
$[\mathrm{tension}]/[\mathrm{length}]$.

Eq.~\eqref{eq:stress-metric-relation} depends on the cell positions $\rvec(\xivec)$, which are physical observables, not mere coordinates\footnote{
Importantly, Eq.~\eqref{eq:stress-metric-relation}, and its consequence Eq.~\eqref{eq:stress_w}, are \emph{not} ``transformation laws'' that relate the stress tensor in different coordinates (akin to Cauchy and Piola-Kirchhoff stress~\cite{Marsden.etal1984}). 
Rather, they describe how the Cauchy stress depends on deformation, like a stress-strain relation in conventional elasticity.}.
Of particular interest are mechanically equilibrated embeddings, which must fulfill force balance:
\begin{align}
    \partial_\alpha \sigma_{\alpha\beta}^\mathrm{tot} = \partial_\alpha \sigma_{\alpha\beta} - \partial_\beta p =0
    \label{eq:div_sigma_tot}
\end{align}
Physically, tensile forces $\partial_\alpha \sigma_{\alpha\beta}$ must balance against pressure gradients $\partial_\beta p$.
From Eqs.~\eqref{eq:stress-metric-relation},~\eqref{eq:div_sigma_tot} we can already anticipate a special role for a configuration $\rvec(\xivec)$ where the tension metric is isotropic. In such an \emph{isothermal configuration}, $\bm{\sigma}$ is isotropic and trivially balanced by pressure, defining an emergent stress-free reference configuration.

Let us briefly summarize the mechanical problem that we have set up so far. Given a tension metric $\mathbf{g}(\xivec)$, a cell density $n(\xivec)$, and an equation of state $P(n)$, we want to find physical configurations $\rvec(\xivec)$ that satisfy
\begin{gather}
    \bm{\sigma}(\rvec) =  p_0 \bm{\epsilon}^T \! \cdot \mathbf{g}(\rvec)^{1/2} \cdot \bm{\epsilon} \tag{SM} \label{eq:SM}\\
    \mathrm{div}_{\rvec} \, \bm{\sigma}(\rvec) - \nabla_{\!\rvec} p(\rvec) = 0 \tag{FB} \label{eq:FB} \\
    p(\rvec) = P\bigl(n(\rvec)\bigr) \tag{EoS} \label{eq:EoS}
\end{gather}
The map $\rvec(\xivec)$ enters these equations through the Jacobian factors that appear in $\mathbf{g}(\rvec)$ and $n(\rvec)$, see Eqs.~\eqref{eq:metric_transform} and~\eqref{eq:pressure_constitutive}. The force balance equation~\eqref{eq:FB} needs to be supplemented by mechanical boundary conditions.
In the following, we introduce a mathematical framework to solve the coupled equations~\eqref{eq:SM}, \eqref{eq:FB}, and~\eqref{eq:EoS} and determine when (stable) force-balanced states exist.
We first study the statics problem to see how tissue shape and response to external forces emerge from fixed $\mathbf{g}(\xivec)$. 
Then, we turn to quasi-static shape dynamics due to changes in $\mathbf{g}(\xivec)$, and finally generalize our framework to describe plastic flow (cell rearrangement).

\subsubsection{Complex coordinates and (quasi)conformal maps}

To solve Eq.~\eqref{eq:div_sigma_tot}, we will use complex coordinates $z=z_1+i z_2$. 
Complex notation converts matrix algebra to scalar algebra and greatly simplifies the description of conformal maps, which will play an important role.
A symmetric matrix $\mathbf{q}$ (e.g., the metric $\mathbf{g}$) defines a quadratic form $\mathbf{v} \mapsto q_{\alpha\beta} v^\alpha v^\beta$ which becomes\footnote{Note: the literature often uses the reverse convention, $q_{zz} \leftrightarrow q_{\bar z\bar z}$.}
\begin{subequations}{\label{eq:complexified_symmetric}}
\begin{align}
    \label{eq:quadratic_form}
    4 q_{\alpha\beta} v^\alpha v^\beta &= (q_{z \bar z}+ q_{\bar z z})v\bar v+ q_{zz} \bar{v} \bar{v}  + q_{\bar{z} \bar{z}}  v v, \\
    \mathrm{where}\;\;
    q_{z\bar z} &= q_{\bar z z} = q_{11} + q_{22} = \mathrm{Tr}\, \mathbf{q}, \nonumber \\
    q_{z z} &= \overline{q_{\bar z \bar z}} = q_{11}-q_{22} + 2iq_{12}
\end{align}
\end{subequations}
Since $q_{zz}$ encodes the deviatoric part $\dev\mathbf q = \mathbf q - \tfrac12(\mathrm{Tr}\,\mathbf q)\,\mathbb I$, it measures the anisotropy.
The \emph{Beltrami coefficient} $\mu_q$ measures the relative anisotropy and is defined by:
\begin{align}
    \mu_q
    = \frac{q_{zz}}{ q_{z\bar z}+ \sqrt{q_{z\bar z}^2 - |q_{zz}|^2} } = \frac{q_{zz}}{\mathrm{Tr}\,\mathbf{q} + 2 \sqrt{\det \mathbf{q}}}
\end{align}

Note that as a \emph{coordinate}, the complex number $z = z_1 + i z_2$ contains the same information as the Cartesian vector $(z_\alpha) = \mathbf{z} = (z_1, z_2)$. We use the two interchangeably. However, it is convenient to write \emph{functions} $f$ as $f(z,\bar z)$ with formally independent variables $z$, $\bar z$.
With this convention, vector calculus can be expressed in terms of the Wirtinger derivatives $\partial_z=(\partial_1-i\partial_2)/2, \; \bar\partial_z=(\partial_1+i\partial_2)/2$. For example, for real-valued potentials $\theta, \psi$, $2\bar\partial_z  \theta$ is the gradient, while $2i\bar\partial_z \psi$ is a solenoidal vector field. 
The Euclidean Laplace operator is $4\partial_z \bar{\partial}_z = \partial_1^2 + \partial_2^2 =: \Delta$.
The Jacobian matrix of a map $f(z,\bar z)$ is denoted $\partial_{\mathbf{z}} f$, with bold-faced subscript:  
\begin{align}
   \partial_{\mathbf{z}} f = \begin{bmatrix} \partial_z f & \bar \partial_z f \\[0.4em]  \partial_z \bar f & \overline{\partial_z f} \end{bmatrix}
   \label{eq:complexified_Jacobia}
\end{align}
Note that the matrix entries in Eq.~\eqref{eq:complexified_Jacobia} are not in Cartesian $z_1,z_2$ but in complexified $z,\bar{z}$ coordinates.

Maps satisfying  the Cauchy--Riemann equation $\bar \partial_z f = 0$, i.e., $z\mapsto z'=f(z)$, are conformal (angle-preserving). Their Jacobian is  a diagonal matrix
$\partial_{\mathbf{z}} f = \mathrm{diag}[\partial_z f, \overline{\partial_z f}]$, combining rotation by the phase $\phi_{z|z'} = \arg \partial_z f$ and scaling by the conformal factor $\lambda_{z|z'} = |\partial_z f|$.
The subscript $z|z'$ indicates the map's source and target. Furthermore, for conformal maps, $\Delta \log \lambda_{z|z'} = 4\partial_z\bar\partial_z \Re[ \log (\partial_z f)] = 0$.
More generally, the magnitude and orientation of a map's shear is measured by the Beltrami coefficient $ \mu_{z|z'} = \bar \partial_z f  / \partial_z f$ which vanishes for a conformal map.
Note that the Beltrami coefficients of the map $f$ and of the quadratic form $(\partial_{\mathbf{z}} f)^\dagger \cdot (\partial_{\mathbf{z}} f)$ agree; $()^\dagger$ being the Hermitian adjoint.
Geometrically, $f$ maps infinitesimal circles to ellipses with semi-axes proportional to $1\pm |\mu_{z|z'}|$. If $|\mu_{z|z'}| < k < 1$ is uniformly bounded, $f$ is called \emph{quasiconformal} (QC).
In particular, QC maps preserve orientation.
Notation-wise, their argument distinguishes QC maps $f(z,\bar{z})$ (also written $f(\mathbf{z})$), from conformal maps $f(z)$. In the following, we use the formalism of QC maps to develop our continuum theory (Ref.~\cite{Cislo.etal2025} used a similar QC approach to describe the growth of 2d tissues).

\subsubsection{Isothermal coordinates}

We now use the mathematics of QC maps to define a convenient parametrization of the tension triangulation.
In 2d, a metric has 3 independent components, while an embedding map has two degrees of freedom. Thus, only a scalar degree of freedom remains. It is therefore always possible to find a set of \emph{isothermal} coordinates~\cite{Lee2018} in which the metric is isotropic, characterized by a single scale factor.
We begin by identifying isothermal coordinates $z(\xi, \bar \xi)$ for the tension metric. Using Eq.~\eqref{eq:complexified_symmetric}, isotropy requires
\begin{align}
    4g_{\alpha\beta} d\xi^\alpha d\xi^\beta &= (\mathrm{Tr}\sqrt{\mathbf{g} })^2 |d\xi+ \mu_g d\bar\xi|^2  \nonumber \\
    &\stackrel{!}{=} 2 g_{z\bar z} dz d\bar z = 2g_{z\bar z} |\partial_\xi z \, d\xi + \bar\partial_\xi z \, d\bar\xi  |^2
\end{align}
where the Beltrami coefficient  $\mu_g =  g_{\xi\xi} / (\mathrm{Tr} \,\mathbf{g}+2\sqrt{\det\mathbf{g}})$ measures the anisotropy of $\mathbf{g}$.
Therefore, $z(\xi, \bar \xi)$ must fulfill a  so-called Beltrami equation:
\begin{align}
  \bar\partial_\xi z(\xi, \bar \xi) = \mu_g \partial_\xi z(\xi, \bar \xi) \quad \Leftrightarrow \quad  \mu_{\xi|z} = \mu_g
  \label{eq:beltrami_g}
\end{align}
The QC map from Lagrangian coordinates $\xi$ to isothermal coordinates $z$ exactly removes the shear part of the metric. (By abuse of notation, we will denote both the coordinates and the map $\xi\mapsto z(\xi,\bar\xi)$ by $z$). 
By the Measurable Riemann Mapping theorem~\cite{Ahlfors1966}, solutions to Eq.~\eqref{eq:beltrami_g} exist as long as $|\mu_g| < k$ for some $k<1$ (microscopically, the tension triangles must be uniformly non-degenerate). This fact will play a crucial role in guaranteeing the existence of mechanically stable states.

\begin{figure}
    \centering
    \includegraphics{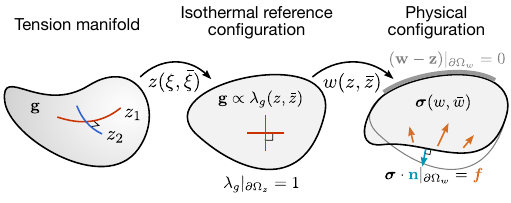}
    \caption{
    Isothermal coordinates ``flatten'' the tension manifold without shear.
    We isothermally map to the ``natural domain'' defined by the boundary condition $\lambda_g|_{\partial \Omega} = 1$.
    The quasiconformal map $w(z,\bar{z})$ maps the reference configuration to the physical domain. It determines the macroscopic stress tensor $\bm{\sigma}$ and is subject to physical boundary conditions (clamped boundary as gray shading, traction forces as red arrows). 
    }
    \label{fig:isothermal}
\end{figure}

The metric thus becomes isotropic and is fully described by its conformal factor $\lambda_g$
\begin{align} \label{eq:g-isothermal}
    g_{z z} =0, \; g_{z \bar z} = 2\lambda_g^2(z, \bar z) \; \Leftrightarrow\; g_{\alpha\beta} = \lambda_g^2 \delta_{\alpha\beta}
\end{align}
Isothermal coordinates are conformal and do not distort angles (see Fig.~\ref{fig:isothermal}; 
Appendix~\ref{app:hemisphere} discusses the example of a hemisphere).
Note that $\xi \mapsto z$ is conformal as a map from the tension manifold with metric $g$ to the Euclidean plane. In the coordinate plane, the map is QC with Beltrami coefficient $\mu_{\xi|z}=\mu_g$.

The Liouville equation determines the Gaussian curvature of an isotropic metric:
\begin{align}
    K_g = -\lambda_g^{-2} \Delta \log \lambda_g
    \label{eq:Gaussian_curvature}
\end{align}
where $\Delta = 4\partial_z {\bar \partial}_z$ is the Euclidean Laplacian.
More generally, the St.~Venant formula~\cite{Muskhelishvili1977} gives the curvature of $g_{\alpha\beta}=\lambda_g^2(\delta_{\alpha\beta} +dg_{\alpha\beta})$ to linear order:
\begin{align}{\label{eq:St-Venant}}
    K_{g} = -\lambda_g^{-2} \big(&\Delta \log \lambda_g + \partial_z\bar\partial_{z} d g_{\bar z z} 
    -\Re[\partial_z^2 d g_{zz}] 
    \big)
\end{align}

Note that the composition of a solution to Eq.~\eqref{eq:beltrami_g} with any conformal map yields another solution (indeed, for conformal maps, $\Delta \log \lambda_{z|z'}=0$, so Eq.~\eqref{eq:Gaussian_curvature} is invariant).
The BCs fix the conformal (``homogeneous'') solution. Throughout, we use ``natural BCs'' 
\begin{align}
    \label{eq:natural_bcs}
    \lambda_g(z,\bar z)|_{\partial\Omega} =  1
\end{align}
where $\partial\Omega$ is the system boundary.
If $\mathbf{g}$ is flat, this BC ensures $g_{z\bar z} = 2$, $\lambda_g = 1$ everywhere (Fig.~\ref{fig:isothermal}).

\subsection{Statics: Emergent elasticity}{\label{sec:statics}}

We now use the complexified formalism to solve Eq.~\eqref{eq:div_sigma_tot}, i.e., identify the tissue's mechanically balanced state and its response to deformations, for a given tension metric and target cell density.
Before getting into the mathematics, we outline the physical picture.

In mechanical balance, tensional stress is balanced by pressure.
We identify a configuration in which the tension metric, and hence the stress, is isotropic and can be \emph{canceled} by pressure. Thus, the tension metric defines an effective, stress-free reference state (Sec.~\ref{sec:reference}). This state exists even if tensions are locally anisotropic.
We then impose the EoS $P(n)$, which links pressure to cell density. As a result, force balance uniquely determines the shape of the tissue. Residual stresses appear if the EoS is not compatible with the reference state pressure.

In response to deformations, the system behaves like an emergent isotropic elastic solid. 
Shear rigidity, the defining property of a solid, is \emph{independent} of the pressure EoS and arises from the active tensions alone.
Effectively, we begin with a system that is a fluid at the microscale, but behaves like a solid at the macroscopic scale. In particular, the emergent shear modulus is always positive. The emergent elastic response is generally nonlinear (Sec.~\ref{sec:iso_conformal_decomp}). By linearizing, we connect to the familiar language of elasticity theory (Sec.~\ref{sec:linearized}). 

Over longer timescales, tensions and cell densities $n$ can change due to biological dynamics. The system is no longer a mere elastic solid, but can ``morph'' due to internal remodeling (Sec.~\ref{sec:dynamics}).
Large deformations lead to yielding due to cell rearrangement (Sec.~\ref{sec:topology_anisotropy}).

\subsubsection{Emergent reference state}{\label{sec:reference}}

We introduced isothermal $z(\xi,\bar{\xi})$ as a convenient parametrization of the tension manifold. We now interpret $z(\xi,\bar{\xi})$ as a specific physical configuration, i.e.\ a map that places the Lagrangian material element $\bm{\xi}$ at physical position $\rvec(\bm{\xi}) = \mathbf{z}(\bm{\xi})$ in the plane.
In the $\mathbf{g}$-isothermal configuration, the tensile stress becomes isotropic. Indeed, by Eq.~\eqref{eq:stress-metric-relation}:
\begin{align}
    \sigma_{zz}=0, \; \sigma_{z\bar z} = 2 p_0 \lambda_g(z, \bar z)
\end{align}
Therefore, with a pressure field
\begin{align}
    \label{eq:pressure_conformal}
    p = p_0 \lambda_g(z, \bar z),
\end{align}
the macroscopic stress can be made to vanish identically: $\sigma_{\alpha\beta}^{\text{tot}} = \sigma_{\alpha\beta} -p\delta_{\alpha\beta} = 0$.
The $\mathbf{g}$-isothermal configuration, thus, is a stress-free reference configuration which \emph{emerges} from a stress-only starting point.
A liquid bubble raft provides a simple physical example. Surface tension balances against pressure so that the raft as a whole is stress-free (cutting it does not lead to macroscopic recoil).
Equations~\eqref{eq:pressure_conformal} and~\eqref{eq:Gaussian_curvature} together imply a Poisson equation for the pressure in the stress-free configuration: 
\begin{align}
    \label{eq:Liouville_pressure}
    \Delta_z \log (p/p_0) = -(p/p_0)^{2}K_g
\end{align}

In a physical configuration, the pressure field $p$ is not an independent variable, but must obey an EoS $p=P(n)$, Eq.~\eqref{eq:pressure_constitutive}.
The stress-free reference state is physically attainable only if
\begin{equation} \label{eq:pressure-density-consistency}
    p_0 \lambda_g(\mathbf{z}) = P(n(\mathbf{z})).
\end{equation}
For a flat tension metric, $\lambda_g = 1$, motivating the definition of a \emph{reference cell density} $n_0$ such that $P(n_0)=p_0$. Equation~\eqref{eq:pressure-density-consistency} then reduces to the compatibility condition
$n(\mathbf{z}) = n(\xivec) /\det[\partial_\xivec\mathbf{z}]= n_0$.
If Eq.~\eqref{eq:pressure-density-consistency} is not fulfilled, the $\mathbf{z}$-configuration is not mechanically balanced. The tissue will adopt a different configuration with residual stresses that remain even when no traction is applied to the boundary.
In this case, we say that the ``tension density'' $\lambda_g(z,\bar z)$ is \emph{incompatible} with the cell density $n(z,\bar z)$. We will see below that this incompatibility acts as a source term for macroscopic stress (a related, but distinct notion of ``incompatibility'' arises in metric elasticity~\cite{Efrati.etal2009}, see Discussion).
Nonetheless, the isothermal $\mathbf{z}$-configuration will be a useful reference for the resulting general force-balanced states.
In particular, it will be the starting point for linearizing the theory.

\subsubsection{General force-balanced states and conformal-isogonal decomposition}\label{sec:iso_conformal_decomp}

To relate the stress-free reference state $\mathbf{z}(\xivec)$ to the realized physical state, we need to calculate how the macroscopic tensile stress $\bm{\sigma}$ changes when the system is deformed from the isothermal reference into a new configuration, $z \mapsto w(z,\bar z)$. Physically, a deformation moves and reorients the local active force dipoles (cell interfaces), thereby modifying the macroscopic stress tensor. Mathematically, $\mathbf{g}$ transforms according to Eq.~\eqref{eq:metric_transform}. Using $g_{ab}(\mathbf{z}) = \lambda_g^2(\mathbf{z}) \delta_{ab}$, Eq.~\eqref{eq:stress-metric-relation} becomes:
\begin{align}
    \label{eq:square_stress_complex}
    \begin{aligned}
     [\bm{\sigma}^2(\mathbf{w})]_{\alpha\beta} &=  p_0^2 \epsilon_{\alpha\gamma} g_{\gamma\delta}(\mathbf{w}) \epsilon_{\beta \delta} \\
     &= p_0^2\lambda_g^2(\mathbf{z})
     \epsilon_{\alpha\gamma}
     \frac{\partial z_\kappa}{\partial w_\gamma} \frac{\partial z_\kappa}
     {\partial w_\delta} \epsilon_{\beta\delta} 
    \end{aligned}
\end{align}
The stress is thus determined by the inverse of the Jacobian $(\partial_{\mathbf{z}} w)$.

Complex notation makes taking the square root of Eq.~\eqref{eq:square_stress_complex} trivial, since
\begin{align}
    \begin{aligned}
         g(\mathbf w)_{\alpha\beta} dw^\alpha dw^\beta&= \left(\lambda_g |\partial_wz|\right)^2 \, \left\vert dw+\mu_{w|z}d\bar w\right\vert^2
    \end{aligned}
\end{align}
Since conjugation by $\epsilon$ flips the sign of the anisotropic part,
so the stress tensor components read
\begin{align}
    \label{eq:stress_w}
    \sigma_{w {\bar w}} = 2p_0 \lambda_g  |\partial_w z|, \quad  \sigma_{w w} =- 2 p_0 \lambda_g  |\partial_w z| \mu_{w|z} 
\end{align}
Eq.~\eqref{eq:stress_w} relates spatial deformation to the changes in physical stress, and can be interpreted as an emergent stress-strain relationship (see the footnote after Eq.~\eqref{eq:stress-metric-relation}).
In particular, we see that deviatoric strain relative to the isothermal reference $\mu_{w|z}$ generates deviatoric stress $\sigma_{w w}$, with an effective positive shear modulus $p_0\lambda_g/2$ (we will return to the effective moduli in familiar Cartesian notation below).
This positivity is a non-trivial result, since the elastic response does not derive from a energy but from a change in the embedding of active force units.

Crucially, the stress anisotropy is \emph{not} identical to the ``microscopic'' tension anisotropy (differences in interface tensions $\tau_{ij}$, or, geometrically, anisotropy of tension triangles).
We will return to this important distinction in Sec.~\ref{sec:topology_anisotropy} where we define a tension anisotropy field in the continuum.

In Cartesian notation, the matrix square-root of Eq.~\eqref{eq:square_stress_complex} can be evaluated via the singular value decomposition of $\partial_{\mathbf{z}}\mathbf{w}=U\cdot \Lambda \cdot V^T$, and the 2d matrix identity $\bm{\epsilon}^T\cdot M^{-1} \cdot \bm{\epsilon} = M^T/\det M$:
\begin{align}
    \label{eq:stress_w_cartesian}
    \bm{\sigma}(\mathbf{w}) 
    = p_0 \lambda_g \frac{U\cdot \Lambda\cdot U^T}{\det\Lambda_{}}
\end{align}
The companion paper~\cite{Claussen.etal2026a} derives the stress-metric relation in the form of Eq.~\eqref{eq:stress_w_cartesian} by explicit coarse-graining.

In mechanical balance, the divergence of the tensile stress must be compensated by a pressure gradient.
In complexified notation, force balance Eq.~\eqref{eq:div_sigma_tot} reads
\begin{align}
\label{eq:balance_w}
    \bar \partial_w \sigma_{{\bar w} w}+ \partial_{w} \sigma_{ w  w} &= 2\bar \partial_{w} p
\end{align}

Applying $\partial_w$ on Eq.~\eqref{eq:balance_w} gives:
\begin{align}
    \partial_{w} \bar \partial_w \sigma_{{\bar w} w } + \partial_{w}^2 \sigma_{w w} &= 2\partial_{w} \bar \partial_{w}p = \tfrac{1}{2} \Delta p
\end{align}
Since $p$ and $\sigma_{{\bar w} w}$ are real, this entails a solvability condition:
\begin{eqnarray}\label{eq:solvability}
\Im \!\left [\partial_{w}^2 \sigma_{w w} \right ]=\Im \!\left[ \partial_{w}^2 \left( \lambda_g \frac{|\partial_w z|}{\partial_w z} \bar \partial_w z \right) \right]=0.
\end{eqnarray}
For a flat tension metric ($K_g = 0$, so $\lambda_g=1$) and small strain of the map $z\mapsto w$, this condition can be simplified. In Cartesian notation:  
\begin{equation}
    \Delta (\epsilon_{\alpha\beta}\partial_\beta w_\alpha(\mathbf{z})) = 0
\end{equation}
The curl of the deformation $w$ must be harmonic. Equation~\eqref{eq:solvability} constrains force-balanced physical configurations $w(z,\bar z)$. Two classes of solutions are readily apparent:

\paragraph{Conformal deformations.} First, conformal maps $z = f(w)$, for which $\bar \partial_w z = 0$, so Eq.~\eqref{eq:solvability} holds trivially.
Under conformal maps, $\sigma$ remains isotropic, while the pressure required for force balance is $p\mapsto p/\lambda_{z|w}$.
This allows broadening the conditions under which a stress-free configuration exists. 
If the reference state $\mathbf{z}(\xivec)$ fails to respect  Eq.~\eqref{eq:pressure-density-consistency}, we can look for a conformal map $w(z)$ so that $P\bigl(n(\mathbf{w})\bigr)=p_0\lambda_g/\lambda_{z|w}$. Due to the restriction $\Delta\log \lambda_{z|w}=0$, this map exists provided that $\Delta \left(B\delta  n-\log\lambda_g\right)=0$ (where $\delta n = n/n_0-1$ and we used the linearized EoS). 
We will confirm this condition below.

As an aside, we note that the relation $p=p_0/\lambda_{z|w}$ between conformal deformations, which curve line elements, and pressure gradients can be understood as a continuum manifestation of the Young--Laplace law~\cite{Weaire.etal2005}. For simplicity, suppose $\lambda_g=1$ and consider a straight line with normal $\mathbf{n}$ in the reference configuration ($z$-coordinates). Through a conformal deformation it acquires a curvature $1/R = \tfrac{2}{|\partial_w z|} \mathbf{n} \cdot \bar\partial_w |\partial_w z|$ (Appendix~\ref{app:curvature_conformal}), while the pressure drop $dp$ across the line element changes to $2p_0 \mathbf{n} \cdot (\bar \partial_w |\partial_ w z|) dr$. Finally, in $w$ coordinates, the tension metric is $d\tau = p_0 \lambda_g |\partial_w z| dr$. Combining, we find $d\tau /R = dp$ -- an exact continuum form of the Young--Laplace law with local line tension $d\tau$.

\paragraph{Isogonal deformations.} Second, for $\lambda_g=1$, a map $z(w,\bar w) = w + 2\bar \partial_w \theta(w,\bar w)$ defined by the gradient of a real potential $\theta$ is a solution since
\begin{align}
    \Im \!\left[ \partial_{w}^2 \left(\tfrac{|\partial_w z|}{ \partial_w z} \bar \partial_w z \right) \right] =  \Im \!\left[ 2\partial_{w}^2 \bar \partial_w^2 \theta \right]= 0 
\end{align}

We refer to such deformations as \emph{isogonal} since they are irrotational and, thus, preserve the angles of cell interfaces in the discrete setting \cite{Noll.etal2017,Claussen.etal2026a}. (Note that this notion of isogonal is different from the term ``isogonal mappings'' which is sometimes used to refer to maps that preserve angle magnitude but not necessarily orientation.)
The corresponding stress tensor reads 
\begin{align} \label{eq:isogonal-stress}
    \sigma_{w\bar w} = 4p_0\,\partial_w \bar\partial_w \theta +2p_0, \; \sigma_{ww} = -4p_0\,\bar\partial_w^2 \theta.
\end{align}
Gradient displacements induce anisotropic stress.
Recall that a balanced stress in 2D can always be represented by an \emph{Airy function}
\begin{align}
   \partial_\alpha\sigma_{\alpha\beta}^\mathrm{tot} = 0 \;\; \Rightarrow\;\; \sigma_{\alpha\beta}^\mathrm{tot} = \epsilon_{\alpha\gamma} \epsilon_{\beta \delta} \partial_\gamma\partial_\delta \chi,
\end{align}
or, written in complex coordinates,
\begin{align}
\label{eq:complex_airy}
    \sigma_{w\bar w}^\mathrm{tot} = 4\,\partial_w \bar\partial_w \chi, \; \sigma_{ww}^\mathrm{tot} = -4\,\bar\partial_w^2 \chi.
\end{align}
Comparing Eq.~\eqref{eq:complex_airy} to Eq.~\eqref{eq:isogonal-stress} shows that 
$p_0 \theta$ is the Airy function for the tensile stress induced by the displacement $\partial_w \theta$.
This implies that the tensile component of the stress is balanced ``on its own''; no pressure gradients emerge (i.e.\ $p(\rvec) = p_0$).

For $K_g\neq 0$, the conformal factor $\lambda_g$ is no longer constant. For a small deformation of the reference state, $z(w,\bar w)=w-u$, the Riemannian gradient of $\theta$ with respect to $\mathbf{g}$, $u=-2\lambda_g^{-2}\bar\partial_w\theta$, is a perturbative solution of Eq.~\eqref{eq:solvability}.
To linear order in $u$, Eq.~\eqref{eq:stress_w} gives $\sigma_{ww}=4p_0\lambda_g\bar\partial_w u$, so Eq.~\eqref{eq:solvability} requires $\lambda_g\bar\partial_w u=\bar\partial_w^2\chi$ for some real Airy function $\chi$.
The displacement $u$ satisfies
\begin{multline} \label{eq:riemannian_gradient_identity}
{-}\lambda_g\bar\partial_w u = 2\lambda_g\bar\partial_w\big(\lambda_g^{-2}\bar\partial_w\theta\big)
    = \\ 2\bar\partial_w^2(\theta/\lambda_g) +2(\theta/\lambda_g)\big[\bar\partial_w^2 \log\lambda_g-(\bar\partial_w \log\lambda_g)^2\big].
\end{multline}
It therefore solves Eq.~\eqref{eq:solvability}, with $\chi=\theta/\lambda_g$, up to the term in square brackets, which can be neglected to first order in conformal factor gradients\footnote{
In a special case, this result can be derived from a Taylor expansion. Suppose the non-constant conformal factor arises from a flat conformal deformation $z'=f(z)$, so  $\lambda_g^{-1} = \lambda_{z|z'} = |\partial_z f|$. n the $z$-configuration, $\lambda_g=1$ and the compatible displacement is $2\bar\partial_z\theta$. In the conformally deformed configuration, the displacement is $u=-2\lambda^{-2}\bar\partial_w\theta$:
\begin{align}
    f(z-2\bar\partial_z\theta) &\approx  f(z) - 2( \partial_z f)(\bar\partial_z\theta) 
    = f(z) - 2\lambda^2 \bar\partial_{f(z)} \theta 
    \nonumber
\end{align}}.
In contrast to the flat case, these modes generally change the pressures (i.e., pressure gradients are required to maintain force balance, Eq.~\eqref{eq:balance_w}).

The manifold of force-balanced states parametrized by $\theta$ corresponds to the \emph{isogonal} ``breathing'' modes of the discrete ATN model for $K_g=0$ \cite{Noll.etal2017,Claussen.etal2026a}, which are closely linked to so-called power diagrams in the mathematical literature on polygonal tilings~\cite{Aurenhammer1987}. 
We therefore refer to $\theta$ as the isogonal potential. Conformal displacements also have an exact equivalent at the cell-level -- a certain ``local'' Möbius symmetry of force-balanced tilings. The companion paper~\cite{Claussen.etal2026a} provides a detailed dictionary between the discrete and continuum theories. 
Conformal and isogonal displacements exhaust all solutions to Eq.~\eqref{eq:solvability} (proven for small strain and flat $\mathbf{g}$ in Appendix~\ref{app:isogonal_conformal_decomp}). Thus, any physical configuration $w$ can be obtained from the reference $z$ by summing a potential and a conformal deformation. As we will see below, these two modes suffice to accommodate arbitrary deformations imposed on the system boundary.

In summary, starting from Lagrangian labels $\xi$, we mapped first to the $\mathbf{g}$-isothermal reference configuration $z$, and then to the physical configuration $w$:
\begin{align}\label{eq:mapping_summary}
    \xi \xmapsto{\mathbf{g}\mathrm{-isothermal}} z(\xi, \bar \xi) \xmapsto{
    \text{force balance}
    } w(z,\bar z).
\end{align}
Appendix~\ref{app:example_flat} provides an elementary example of this mapping.
While force balance in the bulk imposes a conformal--isogonal structure on the deformation $w$,  BCs and a specific constitutive relation for pressure are required to determine the physical configuration completely.

\subsubsection{Freely compressible cells}

By Eq.~\eqref{eq:solvability}, for all isogonal--conformal configurations, there is a pressure field that brings the system into mechanical equilibrium.
However, physically, the pressure field must also obey the EoS $P(n)$, which we impose now.

Let us first discuss a specific case, the continuum equivalent to the ``classic ATN'' setting investigated in Ref.~\cite{Noll.etal2017}.
For a flat tension metric $\mathbf{g}$, Eq.~\eqref{eq:pressure_conformal} implies a constant pressure $p=p_0$.
In the bulk, tensions are in force balance ``on their own'': only at the boundary, a pressure drop $p_0$ is required for normal force balance.
This state can be realized by freely compressible cells with an ``EoS'' $p = p_0$ (i.e., cells can freely exchange volume with one another, and $p_0$ acts as a Lagrange multiplier for the total tissue area).

Purely isogonal deformations of the reference state do not require pressure gradients, since the tensile stress is balanced on its own:
\begin{align}
    \hspace{-.4em}
    \bar \partial_w \sigma_{w\bar w} + \partial_w \sigma_{ww}  = 2p_0 \, [ \bar\partial_w(\partial_w \bar\partial_w \theta) -   \partial_w (\bar \partial_w^2\theta)] = 0
\end{align}
Since the EoS is trivial, $P(n) = p_0$, pressure also remains invariant. Isogonal modes in the bulk are therefore true zero-energy modes.

Under conformal deformations $w(z)$, however, the tensile stress $\sigma_{w\bar w}=2p_0|\partial_ w z|$ becomes non-uniform, implying pressure gradients.
To sustain such pressure gradients, a finite compressibility (a non-constant $P(n)$) is required.
Now, force balance $\partial_\beta p =\partial_\alpha \sigma_{\alpha\beta}$ selects a single physical solution among the isogonal--conformal family of solutions by fixing the isogonal potential $\theta$ in the bulk. Mechanical BCs further fix the boundary DOFs of the conformal and isogonal maps.

\subsubsection{Linearized theory: effective Cauchy elasticity}{\label{sec:linearized}}

To understand the physics resulting from the interplay of conformal--isogonal structure and the constitutive law for pressure, we now expand stress and pressure for small deformations $w= z+u(z,\bar z), \; |\bar\partial_zu|,|\partial_z u| \ll 1$ around the isothermal, stress-free reference state. We further assume small cell-density deviations in the reference state, $n(z,\bar z) = n_0(1+ \delta n(z,\bar z))$, and
small curvature $K_g$ (hence, $\log \lambda_g$ is also small). In the discrete setting, cell areas are close to the Voronoi areas of the tension triangulation.
Below, we will identify a complete set of linearized equations to be solved for $u$ which determine the physical shape of the tissue.

\paragraph{Linearized stress-strain relationship.}

Near the reference cell density $n_0$, one can linearize the pressure EoS 
\begin{align}
    \label{eq:constitutive_linear}
    P(n)\approx p_0 + p_ 0 B \, \frac{n - n_0}{n_0}
\end{align}
where $B$ is the non-dimensionalized cellular bulk modulus. Evaluating at $n(w,\bar w)$, one obtains
\begin{align}
    \begin{aligned}
    p(w,\bar w) &= P\left(\!\frac{n(z, \bar z)}{\det \partial_{\mathbf{z}} \mathbf{w}} \! \right) \\
    &= p_0 + p_0 B (\delta n- 2\Re[\partial_z u])
    \end{aligned}
\end{align}
Also expanding the tensile stress Eq.~\eqref{eq:stress_w} to linear order in $u$, the total stress becomes
\begin{align}
    \label{eq:stress_w_linear}
    \sigma^\mathrm{tot}_{w\bar w} &= 2p_0 \big\{(2B-1)\Re[\partial_z  u] +  \log\lambda_g -  B \delta n \big\} \nonumber \\
    \sigma^\mathrm{tot}_{ w  w} &= 2p_0 \bar \partial_z  u
\end{align}
In Cartesian notation, Eq.~\eqref{eq:stress_w_cartesian} leads to
\begin{align}
    \label{eq:effective_hooke}
    \begin{aligned}
    \bm{\sigma}^\mathrm{tot}  &= p_0 \big[\dev \bm{\varepsilon} + (B-\tfrac{1}{2})(\mathrm{Tr}\,\bm{\varepsilon})\,\mathbb I+(\log\lambda_g-B\delta n)\,\mathbb I\big],
    \\
    \varepsilon_{\alpha\beta} &= \tfrac{1}{2}(\partial_\alpha u_\beta +\partial_\beta u_\alpha)
    \end{aligned}
\end{align}
where $\bm{\varepsilon}$ is the linearized strain tensor.
Eq.~\eqref{eq:effective_hooke} is mathematically equivalent to linear isotropic elasticity~\cite{Landau.Lifshitz1986} with an isotropic ``prestress'' $p_0(\log\lambda_g -  B \delta n ) \delta_{\alpha\beta}$.
However, the underlying physical picture is quite different: a conventional elastic material is akin to a Hookean spring, where strain relative to the rest length triggers a restoring force.
In an active solid, active force dipoles, whose tension is independent of elastic strain, must be arranged (embedded) to balance against the isotropic pressure.
The map to linear elasticity shows that a force-balanced embedding exists for any active tension configuration $\mathbf{g}$ (with a caveat discussed next). This result relies on the existence of isothermal coordinates $z(\xi,\bar\xi)$, in which the tensile stress is isotropic, and of two compatible modes (isogonal and conformal), which can accommodate elastic deformations in response to external and body forces.

\paragraph{Effective elastic moduli.}

One identifies the effective shear and bulk moduli as $\frac{1}{2}p_0$ and $(B - 1/2)p_0$.
Importantly, the non-zero shear modulus of the system is completely \emph{independent} of the EoS $P(n)$ and the cell-level bulk modulus $B$; it arises from the active tension scale $p_0$ alone. The factor $1/2$ in the moduli comes from linearization of the square-root in the stress-metric relation.
Also note that the elastic moduli are \emph{isotropic}, even though tensions can be locally anisotropic.
% \nc{Remark removed, we have the full, non-linear moduli above.}
The finite shear modulus indicates that active tension networks are effective solids, rather than fluids. 
Furthermore, for $B > 1/2$, both effective moduli are positive. If $B < 1/2$, the total bulk modulus is negative: intracellular pressure is insufficient to balance the isotropic tensile stress. Mechanical stability then requires applying traction forces at the boundary to prevent tissue collapse.
The constitutive relation $P(n)$ thus ensures stable mechanical equilibrium.
If pressures are instead upgraded to independent, active degrees of freedom (like the tensions $\tau_{ij}$), mechanical stability is no longer guaranteed. For example, fixing both pressure and surface tension in a soap bubble renders it unstable.

Thus, up to the overall scale $p_0$, the theory has only one free parameter: the cellular bulk modulus $B$. Let us briefly discuss four special cases:
(i)~For incompressible cells, $B\rightarrow\infty$ forcing $\partial_\alpha u_\alpha = \delta n$.
(ii)~If pressures follow the ideal gas law $p(n) =p_0 n/n_0$, i.e.\ $B=1$, the \emph{total} bulk and shear moduli are equal, which may be relevant to experiments on conventional foams with air-filled bubbles.
(iii)~For $B=1/2$, the total bulk modulus vanishes, so that the conformal mode becomes soft (conformal dilations cost no energy).
(iv)~For fully compressible cells $B=0$, the isogonal potential $\theta$ is an unconstrained soft mode since $\bm{\sigma}^\mathrm{tot}$ is balanced for any $\theta$. However, conformal modes, which require pressure gradients, cannot be accommodated, and, as we will see next in Eq.~\eqref{eq:Airy_linear}, no mechanically balanced state exists if $K_g\neq 0$.

\paragraph{Biharmonic equation and effective Airy function.}

By Eq.~\eqref{eq:effective_hooke}, force balance $\partial_\alpha \sigma^\mathrm{tot}_{\alpha\beta}=0$ becomes: 
\begin{align}
    \Delta u_\beta + \left(2B - 1\right) \partial_\beta &(\partial_\alpha u_\alpha) = 2\partial_\beta \left( B\delta n - \log\lambda_g \right)
    \label{eq:navier}
\end{align}
As expected from the solvability analysis above, the general solution to Eq.~\eqref{eq:navier} can be written as $u_\alpha= \partial_\alpha \theta+ \nu_\alpha$ where $\nu_\alpha$ is a conformal vector field~\cite{Stickforth1975,Muskhelishvili1977}. Eq.~\eqref{eq:navier} reduces to:
\begin{align}
    \Delta \theta = -\big(1-(2B)^{-1}\big)\;\partial_\alpha \nu_\alpha + \delta n - B^{-1}\log \lambda_g + C
    \label{eq:PN_condition}
\end{align}
where $C$ is an integration constant which can be absorbed into a uniform scaling.
Due to Eqs.~\eqref{eq:Gaussian_curvature} and~\eqref{eq:PN_condition}, $\theta$ obeys a biharmonic equation:
\begin{align}
    \label{eq:Airy_linear}
    \Delta^2\theta = \Delta \delta n + B^{-1} K_g
\end{align}

In a force-balanced state, the isogonal potential sets the Airy stress function, $\chi=-p_0\theta$.
Using Eq.~\eqref{eq:PN_condition} and the Cauchy--Riemann equations for $\nu_\alpha$, the stress reads:
\begin{align}
    \sigma_{\alpha\beta}^\mathrm{tot} = p_0\left( \partial_\alpha \partial_\beta\theta - \Delta \theta \delta_{\alpha\beta} \right) = -p_0 \, \epsilon_{\alpha\gamma}\epsilon_{\beta\delta} \partial_\gamma\partial_\delta \theta
    \label{eq:Airy_stress_continuum}
\end{align}
That is, the terms involving the conformal mode cancel.
The conformal mode instead defines the internal pressure relative to the stress-free reference:
\begin{equation}\label{eq:excess_pressure}
    p = p_0\lambda_g + \tilde p, \qquad \tilde p = -\tfrac{p_0}{2}\,\partial_\alpha \nu_\alpha .
\end{equation}

Eq.~\eqref{eq:Airy_linear} is analogous to the biharmonic equation for the Airy function in classical elasticity. However, Gaussian curvature of the tension metric $K_g$ and cell density deviations $\delta n$ appear as source terms, implying residual stresses even for traction-free BCs. These stresses result from an incompatibility between a constant cell density and an isotropic tensile stress (the macroscopically stress-free state does not have uniform pressure).
Eq.~\eqref{eq:Airy_linear}  can be solved numerically, or via complex variable techniques (the Kolosov-Muskhelishvili formalism~\cite{Muskhelishvili1977}).

\paragraph{Boundary conditions.}
Together with two BCs on stress $\sigma_{\alpha\beta}$ or displacement $u_\alpha$, the biharmonic equation Eq.~\eqref{eq:Airy_linear} fully determines the tissue shape (Appendix~\ref{app:examples} presents two worked examples).
For example, consider applying a traction force $\mathbf{f}(s)$ to the boundary  $\partial\Omega$, parametrized by arc length $s$ and unit normal $\hat{\mathbf n}(s)$. The stress at the boundary must balance the traction force:
\begin{align}
     \mathbf{f}(s) \stackrel{!}{=} \bm\sigma^{\rm tot}(s)\cdot\hat{\mathbf n}(s)=-p_0\,\partial_s(\bm\epsilon\cdot\nabla\theta)
\end{align}
Writing $\partial_n$ for the normal derivative at $\partial\Omega$, one has:
\begin{equation}\label{eq:traction_bc}
\begin{gathered}
    \partial_s\theta = -\frac{\hat{\mathbf n}\cdot\mathbf F(s)}{p_0}, \quad
    \partial_n\theta = -\frac{(\bm{\epsilon}\cdot \hat{\mathbf n})\cdot\mathbf F(s)}{p_0}, \\
    \mathbf{F}(s) := \int_0^s \mathbf{f}(s')ds'
\end{gathered}
\end{equation}
Traction-free boundaries thus imply $\theta = \partial_n\theta = 0$ on $\partial\Omega$.

Eq.~\eqref{eq:Airy_linear} can also be factorized into two Poisson equations. 
\begin{subequations}\label{eq:factored}
\begin{align}
    \Delta \tilde p &= 0, \label{eq:factored_a}\\
    \Delta\theta &= \left(\delta n - B^{-1}\log\lambda_g \right) + \big(2-B^{-1}\big)\,\tilde p/p_0. \label{eq:factored_b}
\end{align}
\end{subequations}
where $\lambda_g$ is the solution of Eq.~\eqref{eq:Gaussian_curvature}.
Because $\nu_\alpha$ is conformal, it is determined by $\partial_\alpha \nu_\alpha =-2 \tilde p /p_0$ up to a rigid motion (a conformal displacement thus admits a single boundary condition, App.~\ref{app:boundary_conditions}).
The boundary values of $\theta$ and $\tilde p$ are fixed by the two mechanical BCs\footnote{Per Eq.~\eqref{eq:traction_bc}, Dirichlet BCs on $\theta$ encode the normal traction force. The boundary values of $\tilde p$ adjust $\partial_n \theta$ to match the tangential traction force.}. Thus, both isogonal and conformal modes are required to accommodate general boundary forces or displacements.

In summary, boundary forces deform the tissue away from the emergent reference state $z(\xi,\bar\xi)$. Incompatible cell densities and tension curvature lead to residual stresses, present even for free boundaries. The physical stress is governed by effective linear elasticity.
A remarkable consequence is that the shape and macroscopic mechanical properties of the tissue are determined by the tension metric and cell density alone. There are many microscopic tension configurations (tension triangulations) that have the same $\mathbf{g}$ and $n$. On a macroscopic level, they are all equivalent -- the microscopic details are irrelevant for the static problem.

\subsection{Active tension dynamics: Quasiconformal flow}{\label{sec:dynamics}}

We have seen how the cell positions $\xi \mapsto w\big(\xi, \bar\xi \big)$ embed the tension metric into physical space, and how they are determined by force balance.
We next ask how the physical configuration $w(\xi, {\bar \xi},t)$  changes  in response to changes of active tension $\partial_t \mathbf{g}$.
This provides a means for biology to change tissue shape using local internal forces and generate ``tissue flow''~\cite{Streichan.etal2018,Saadaoui.etal2020}.
The tension dynamics are subject to biological control, which involves genetic programs and mechanochemical feedback loops~\cite{Irvine.Wieschaus1994,Heisenberg.Bellaiche2013,Lefebvre.etal2023a,Caldarelli.etal2024}. We consider general active tension dynamics
\begin{align} \label{eq:m_definition}
    \partial_t \log \mathbf{g}(\bm{\xi},t)= \mathbf{g}^{-1} \!\cdot  \partial_t \mathbf{g} = \mathbf{m}(\bm{\xi},t).
\end{align}
The ``tension control'' tensor $\mathbf{m}$ represents dynamic control of tension (e.g., myosin recruitment~\cite{Bertet.etal2004}), which we do not specify further here.
The tension dynamics depend on the biology of the specific system; physics then determines, quite generally, how the tension metric translates into the force-balanced physical configuration.

The tensor $\mathbf{m}$ acts as an effective forcing term that internally drives the morphing of the active solid from one shape to another.
It acts similarly to the ``growth tensor'' in morphoelasticity, a theory of, e.g., plant growth~\cite{Goriely.Moulton2011}.
We emphasize that $\mathbf{m}$ is a specified rate of change of active tension, not a constitutive relation.
In practice, one must provide a concrete \emph{model} for $\mathbf{m}$ based on hypothesized biological mechanisms. 
Note that in addition to the tensions, the target cell density $n_0$, which determines the local pressure, can also evolve adiabatically, for example due to cell growth. We will address these effects in Sec.~\ref{sec:adiabatic_w}.

We now compute how the tension control tensor changes the physical tissue configuration. Sec.~\ref{sec:linear_active_solid} show that each tension increment acts as an effective active stress on top of the emergent elastic solid that arises from the current tension configuration.

\subsubsection{Adiabatic approximation}

We will assume rapid relaxation towards mechanical equilibrium. Consider relaxation dynamics of the cell positions $\rvec$ 
\begin{align}
    \label{eq:relaxation}
    \Gamma \partial_t r_\beta (\xi, {\bar \xi},t)=\partial_\alpha \sigma_{\alpha\beta}(t)  -\partial_{\beta}p(t)
\end{align}
where $\Gamma$ is a friction coefficient. The stress $\sigma$ depends both on $\rvec(t)$ and $\mathbf{g}(t)$ by Eq.~\eqref{eq:stress-metric-relation}, and is thus time-dependent. The relaxation timescale is given by $\Gamma L^2/p_0$, where $L$ is the system length scale. 
In the limit where relaxation is rapid (compared to tension dynamics $\partial_t\log \mathbf{g}$), Eq.~\eqref{eq:relaxation} reduces to force balance:
\begin{align}
    \label{eq:relaxation_balance}
    \partial_\alpha \sigma_{\alpha\beta}(t)  -\partial_{\beta}p(t) \approx 0
\end{align}
Due to rapid relaxation, transient friction or viscous stress will play no important role.
The cells, therefore, adopt the instantaneous equilibrium configuration $\mathbf{w}[\mathbf{g}(t)](\xi,\bar\xi)$, considered as a functional of the time-dependent tension metric $\mathbf{g}(t)$.

In the adiabatic approximation, the tissue flow due to tension dynamics can be computed by the chain rule:
\begin{align}
      \partial_t w[\mathbf{g}(t)](\xi, {\bar \xi}) &=
      \frac{\delta w[\mathbf{g}]}{\delta \mathbf{g}} \circ \partial_t \mathbf{g} =
      \frac{\delta w[\mathbf{g}]}{\delta \mathbf{g}} \circ \left(\mathbf{g}\cdot
      \mathbf{m}\right)(\xi,\bar \xi) \nonumber \\
      & := 
      \int \frac{\delta w[\mathbf{g}](\xi,\bar \xi) }{\delta \mathbf{g}(\xi',\bar\xi') }  \, \left(\mathbf{g}\cdot \mathbf{m}\right)(\xi', \bar\xi') d\xi' d\bar\xi'
      \label{eq:functional_derivative}
\end{align}
The functional derivative $\delta w/\delta\mathbf{g}$ is a convolution-like integral operator, defined on the second line. The adiabatic flow is thus ``non-local'': a local change in contractility induces strain across the tissue.
We evaluate Eq.~\eqref{eq:functional_derivative} by the composition of incremental maps and further 
decompose $w(\xi,\bar\xi)$ into $\xi \mapsto z(\xi,\bar\xi)\mapsto w(z,\bar z)$, from $\xi$ to the reference $z$, and from $z$ to the physical state. 
By the chain rule:
\begin{align}
    \frac{\delta w [\mathbf{g}](\xi,\bar \xi) }{\delta \mathbf{g}} = \frac{\delta w[\mathbf{g}]}{\delta z} \circ \frac{\delta z [\mathbf{g}]}{\delta \mathbf{g}} + \frac{\delta w[\mathbf{g}]}{\delta \bar z} \circ \frac{\delta \bar z [\mathbf{g}]}{\delta \mathbf{g}}
\end{align}
Here, $\delta z/\delta\mathbf{g}$ describes the variation of the isothermal embedding of $\mathbf{g}$, while $\delta w/\delta z$ involves solving an elasticity problem, under the constraints imposed by the physical BCs (e.g., zero normal stress).

\subsubsection{Adiabatic flow of the reference configuration}\label{sec:adiabatic_z}

The dynamics of $z(\xi,\bar\xi, t)$ is given by the Beltrami Eq.~\eqref{eq:beltrami_g}.
Consider an infinitesimal increment
\begin{align}
    \label{eq:metric_change}
    \mathbf{g}(\xivec, t) \mapsto \mathbf{g}(\xivec,t) + \partial_t \mathbf{g}(\xivec, t) \, dt
\end{align}
The resulting increment $z \mapsto z + \dot{z}\,dt$ must reestablish a state of isotropic stress. 
Rather than in Lagrangian $\xi$-coordinates, we express this adiabatic increment in the current $z(t)$ coordinates, so that $\dot{z}\bigl(z(t),\overline{z(t)}, t\bigr)$ is the Eulerian velocity field. For brevity, we write $z = z(t)$. 
Expanding the Beltrami Eq.~\eqref{eq:beltrami_g} in $dt$, and using the definition Eq.~\eqref{eq:m_definition} of the $\mathbf{m}$-tensor leads to the simple PDE
\begin{align}
    \label{eq:beltrami_flow_Eulerian}
    \begin{aligned}
    \bar\partial_z \dot{z}(z,\bar z, t) &= \tfrac{1}{4} m_{zz}(z, \bar z, t), \\\Re[\partial_z \dot{z}]|_{\partial \Omega} &= \tfrac{1}{4}  m_{z\bar z}|_{\partial \Omega}
    \end{aligned}
\end{align}
for the infinitesimal increment of the isothermal coordinates $z$.
The BCs ensure that, throughout the flow, $\lambda_g|_{\partial \Omega} = 1$; i.e., the flow introduces the minimum scale deformation required.
Chaining together a sequence of such infinitesimal QC deformations defines a QC flow (also called a Beltrami flow). Notably, the QC flow equation is linear and can be solved by Green's functions, even for large deformations~\cite{Ahlfors1966}. 
Physically, Eq.~\eqref{eq:beltrami_flow_Eulerian} describes how an anisotropic \emph{change} of active tension induces a corresponding shear deformation of the stress-free reference state.
For traction-free boundaries (and under an additional condition explained below), flow of the reference state $z$ translates directly into flow of the physical configuration $w$.

Equation~\eqref{eq:beltrami_flow_Eulerian} can be brought into a more familiar form by applying $\partial_z$ and changing to Cartesian notation
\begin{align}\label{eq:effective_Stokes}
    \Delta \dot{\mathbf{z}} &= \nabla \cdot  \dev \mathbf{m}
\end{align}
Eq.~\eqref{eq:effective_Stokes} resembles a Stokes flow, driven by the divergence of $\dev \mathbf{m}$ (unlike an actual Stokes flow, there is no pressure or incompressibility constraint). Recall, moreover, that $\mathbf{m}$ is not an active stress but rather the \emph{rate of change} of the tension metric. In the adiabatic regime, the system follows the force-balanced configuration imposed by the internal tension configuration.
The flow rate is thus set by the timescale on which cells regulate contractility, and not by the (rapid) rate of mechanical relaxation. Adiabatic flow is not stress relaxation:
In the $z$-configuration, the total macroscopic stress $\bm{\sigma}^\mathrm{tot}=0$ throughout the adiabatic QC flow.

\subsubsection{Adiabatic flow of the physical configuration}\label{sec:adiabatic_w}

We are ultimately interested in the dynamics of the physical configuration $w$, which may differ from $z$ due to external or body forces. 
By the chain rule, the  physical velocity of a material element $\xi$ is 
\begin{align}
    \tfrac{d}{dt}\mathbf{w}\big(\mathbf{z}(\bm{\xi},t),t\big)= (\partial_\mathbf{z} \mathbf{w}) \cdot \dot{\mathbf{z}} + \dot{\mathbf{w}}. %\nc{I think it's nice to have this one as a full-line equation}
\end{align}
For simplicity, we work to linear order in the strain, as in Sec.~\ref{sec:linearized}, so $\partial_\mathbf{z}\mathbf{w} \approx \mathbb{I}$.
Thus, the Lagrangian cell velocity is $\dot{z} +\dot{w}$; $\dot{w}$ is the passive flow relative to the moving reference $z$.
When integrating over longer times (large changes in $z$), the Jacobian $\partial_\mathbf{z} \mathbf{w}$ must be taken into account.

The flow $\dot w$ is determined by the BCs and the constitutive relation $P(n)$ (see Sec.~\ref{sec:linearized}).
For instance, in the incompressible tissue limit $B \to \infty$, one obtains:
\begin{subequations}\label{eq:flow_w}
\begin{align}
    \Re[\partial_z \! \left(\dot{z} + \dot{w}\right)] &= 0, & \big(\partial_\alpha(\dot{z}_\alpha + \dot{w}_\alpha) &= 0\big) \label{eq:flow_w_a} \\
    \partial_z\bar\partial_z \Im[
    \partial_z \dot{w}] &= 0  & \big(\Delta(\epsilon_{\alpha\beta}\partial_\alpha \dot{w}_\beta) &= 0\big) \label{eq:flow_w_b}
\end{align}
\end{subequations}
\noindent Eq.~\eqref{eq:flow_w_a} enforces incompressibility. 
Eq.~\eqref{eq:flow_w_b} derives from the solvability condition Eq.~\eqref{eq:solvability}, expanded for $|\lambda_g-1| \ll 1$ and small strain: the physical flow \emph{relative} to the changing reference configuration must obey the force-balance compatibility condition.

Let us consider the important special case $m_{z\bar z} =0$ (purely anisotropic tension change), and free BCs.
In this case, $\dot{w}=0$ is a solution of Eq.~\eqref{eq:flow_w_b}. For $\dot{w}=0$ to solve Eq.~\eqref{eq:flow_w_a}, $\dot z$ must be incompressible, i.e., $\dot{z} = 2i\bar\partial_z \psi$ for a real-valued solenoidal potential $\psi$.
Plugging this ansatz into Eq.~\eqref{eq:beltrami_flow_Eulerian} leads to the condition $\Re[\partial_z^2m_{zz}]=0$.
Observe that Eq.~\eqref{eq:St-Venant} implies $\dot{K}_g  =0$. Thus, $\dot{w} = 0$ is the solution to Eqs.~\eqref{eq:beltrami_flow_Eulerian} and~\eqref{eq:flow_w} in the incompressible limit $B \to \infty$ when the rate of change of $\mathbf{g}$ is flat and trace-free. In this special case, ``reference flow'' translates directly into physical flow and the system remains macroscopically stress-free.

\paragraph{Effects of cell growth}

In addition to tuning tensions, tissues can also adiabatically modify the target cell density $n_0$, e.g., by growth or apical area expansion.
In the EoS Eq.~\eqref{eq:constitutive_linear}, this enters as a slowly varying target density $n_0e^{-\eta(\xivec,t)}$, so that $P\approx p_0[1+B(n\,e^{\eta}/n_0-1)]$.
The reference configuration $z(\xi,\bar\xi)$ only depends on the tension metric and is unaffected, but the physical equilibrium state changes ($\delta n\to\delta n+\eta$ in the linear theory of Sec.~\ref{sec:linearized}).
For incompressible tissues, the result is particularly simple. The divergence of the physical flow must match the growth rate:
\begin{align}
    \Re[\partial_z \!\left(\dot{z} + \dot{w}\right)] = \partial_t\eta
\end{align}

\subsubsection{Tension increment as active stress on an active solid}\label{sec:linear_active_solid}

In general, one must solve the elasticity problem Eq.~\eqref{eq:flow_w} to determine $w$ in terms of $z$.
In fact, Eqs.~\eqref{eq:effective_Stokes} and~\eqref{eq:flow_w} together are equivalent to the equations for a small displacement field $u = \dot{z} + \dot{w}$ induced by a stress $p_0 \bm{\epsilon}^T\cdot \mathbf{m} \cdot \bm{\epsilon}/2$ on an incompressible linear elastic background (the rotation matrix $\bm{\epsilon}$ and the factor $1/2$ arise from the stress-metric relation).
The dynamical tension increment acts like an active stress, establishing a bridge between the tension-metric formalism and the existing active matter literature.
Note that here, the elastic background is \emph{itself} an emergent property, defined by the active forces at the preceding time point.

For simplicity, we assume that the effective elastic background is initially stress-free (cf. Sec.~\ref{sec:linearized}), and the pressure is $p=p_0$.
We thus consider an incompressible, linear elastic solid with a shear modulus $p_0/2$, and an active stress $p_0 \bm{\epsilon}^T\cdot \mathbf{m} \cdot \bm{\epsilon}/2$. 
In the incompressible limit, only $\dev \mathbf{m}$ is relevant, and $\bm{\epsilon}^T\cdot \dev \mathbf{m} \cdot \bm{\epsilon} = -\dev \mathbf{m}$.
Thus, the displacement field $\mathbf{u}$ obeys
\begin{equation} \label{eq:active-solid}
    \tfrac{1}{2} \Delta \mathbf{u} - \nabla \tilde p = - \nabla \cdot \mathbf{m}, \quad \nabla \cdot \mathbf{u} = 0.
\end{equation}
where the (dimensionless) pressure $\tilde p$ is the Lagrange multiplier to enforce incompressibility.
To show the equivalence of Eq.~\eqref{eq:active-solid} to Eqs.~\eqref{eq:effective_Stokes} and~\eqref{eq:flow_w}, we set $\mathbf{u} = \dot{\mathbf{z}} + \dot{\mathbf{w}}$. Eq.~\eqref{eq:flow_w_a} becomes $\nabla\cdot\mathbf{u} = 0$ and   Eq.~\eqref{eq:flow_w_b} becomes $\Delta \epsilon_{\alpha\beta} \partial_\alpha (u_\beta - \dot{z}_\beta) = 0$. Using Eq.~\eqref{eq:effective_Stokes} yields
\begin{align}
    \epsilon_{\alpha\beta} \partial_\alpha (\Delta u_\beta - \partial_\gamma m_{\gamma\beta}) = 0
\end{align}
which is the curl of Eq.~\eqref{eq:active-solid}.

Taking the divergence of Eq.~\eqref{eq:active-solid} gives $\Delta \tilde p = \partial_\alpha\partial_\beta m_{\alpha\beta}$. This has the trivial solution $\tilde p = 0$ if $\partial_\alpha\partial_\beta m_{\alpha\beta} = 0$. 
If $\mathbf{m}$ is also purely deviatoric, %(or more generally, $\Delta \mathrm{Tr} [\mathbf{m}] = 0$),
then there exists a vector field $\mathbf{v}$ such that $m_{\alpha\beta} = \partial_\alpha v_\beta + \partial_\beta v_\alpha$. Since the total stress of the active solid is given by $\sigma^\mathrm{tot}_{\alpha\beta} = \tfrac{p_0}{2} (\partial_\alpha u_\beta + \partial_\beta u_\alpha) -\tilde p \delta_{\alpha\beta}- \tfrac{p_0}{2} m_{\alpha\beta}$, this implies the existence of a stress-free solution $\mathbf{u} = \mathbf{v}$. 
This corresponds exactly to the case $u = \dot{z}$, $\dot{w} = 0$ in our framework. Indeed, the conditions $\Re[\partial^2_z m_{zz}] = 0$, $\Delta m_{z\bar{z}} = 0$ correspond exactly to the ``compatibility condition'' $ \partial_\alpha\partial_\beta m_{\alpha\beta}=0$.

In summary, Eqs.~\eqref{eq:beltrami_flow_Eulerian} and~\eqref{eq:flow_w} describe the reestablishment of force balance after a change of stress. At this point, it would appear that a large shape change requires building up increasing tension anisotropy (i.e., tension triangle anisotropy). As we will see next, however, cell rearrangements plastically lock in past tension dynamics while resetting the built-up tension anisotropy. Macroscopically, these dynamics appear as a steady plastic flow.

\subsection{Topological rearrangement}{\label{sec:topology_anisotropy}}

Large tissue deformations cause cells to rearrange, leading to plastic shape changes.
In the discrete model, such T1 processes\footnote{``T1'' stands for ``topological 1''; here, we use ``topology'' to refer to the cell adjacency graph and not the gross topology of the tension manifold, which we assume to be that of a disk.} correspond to edge flips that change the cell adjacency graph (Fig.~\ref{fig:T1}). These flips occur when the length of a cell-cell interface vanishes. The challenge is to represent the inherently discrete T1 processes in the continuum.
In this Section, we introduce an additional degree of freedom, the \emph{adjacency metric}, via which we describe microscale topology as well as cell-shape and tension anisotropy. We then introduce a convenient mathematical framework based on quasiconformal maps in Secs.~\ref{sec:anisotropies_beltrami} and~\ref{sec:kinematics}, and return to the physical consequences of T1s for plastic flow in Secs.~\ref{sec:T1_threshold} and~\ref{sec:T1_relax}.

\subsubsection{Continuum representation of microscale topology, cell shape, and tension anisotropy}\label{sec:three_metrics}

\begin{figure*}
    \centering
    \includegraphics{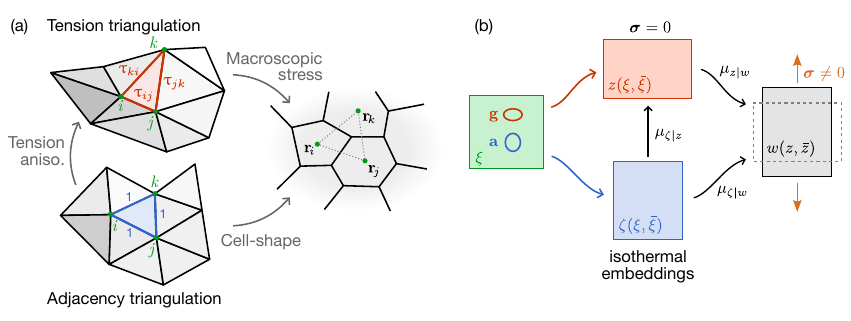}
    \caption{
    (a)~Microscopic geometry with tension triangulation and adjacency triangulation and the corresponding cell tiling. The adjacency triangulation serves as an equilateral reference. The Jacobians of maps between these objects define tension, stress, and cell-shape anisotropy. 
    (b)~Continuum equivalents. Discrete cell labels $i,j,...$ correspond to Lagrangian coordinates $\xi$. Tension and adjacency triangulations define metrics $\mathbf{g}, \mathbf{a}$ that are conveniently represented through their isothermal embeddings (red and blue rectangles). The isothermal embedding of the tension metric defines the stress-free reference state. Geometric incompatibility and boundary traction (orange arrows) deform this into a stressed physical configuration (gray rectangle). The Beltrami coefficients of maps between the embeddings define tension, stress, and cell-shape anisotropy in the continuum.
    }
    \label{fig:maps}
\end{figure*}

Importantly, T1 transitions do not manifest themselves in the tension metric; their description requires an additional degree of freedom.
This is most visible at the microscopic level. Consider a tension manifold that is a subset $\Omega$ of the flat Euclidean plane. There are many different tension triangulations of $\Omega$; but the tension metric (in this case, the Euclidean distance between material points in $\Omega$) \textit{remains the same}. However, re-triangulation 
changes the topology of the cell tiling, i.e., the cell adjacency matrix. 
Thereby, it can change the tension anisotropy (equilateral vs anisotropic tension triangles), the cell density (large vs small tension triangles), and the cell shape.

To capture this information, we introduce a second \emph{adjacency metric} $a_{\alpha\beta}(\bm{\xi})$. Note that $\mathbf{a}(\bm{\xi})$ and the tension metric $\mathbf{g}(\bm{\xi})$ define two \emph{distinct} metrics on the same material manifold, parametrized by Lagrangian coordinates $\bm{\xi}$ attached to material elements (e.g.\ cells).
On the microscopic level, $\mathbf{a}$ is defined so that adjacent cells have unit distance. It thus encodes the cell adjacency graph (the local neighbor relationships).
The adjacency metric ``geometrizes'' this topological information: the adjacency graph is a triangulation where every edge length is set to one (Fig.~\ref{fig:maps}a). 
Cell rearrangements correspond to dynamics of the adjacency metric, since edge flips change the adjacency distance between cells involved in a T1 process (Fig.~\ref{fig:T1}).

We now coarse-grain the discrete adjacency triangulation into a smooth surface.
This discards microscopic detail; several microscopic adjacency triangulations become (approximately) equivalent after smoothing. 
The coarse-grained adjacency metric $\mathbf{a}$ measures the number of ``microscopic'' cells traversed along the displacement $d\bm{\xi}$ between two nearby material elements:
\begin{align}
    dn^2 = a_{\alpha\beta} d\xi_\alpha d\xi_\beta
\end{align}
We dimensionalize $\mathbf{a} \mapsto \mathbf{a}/n_0$ so that $\mathbf{a}$ has units of area\footnote{Microscopically, the cell area in an equilateral triangulation is proportional to the coordination number. Hence, strictly speaking, $\mathbf{a} \mapsto \mathbf{a}/n_0$ only holds if there is no large-scale coordination number gradient.}.
Thus, the isotropic part of $a_{\alpha\beta}$ is the cell density $n = n_0\sqrt{\det \mathbf{a}}$, and the anisotropic part is the cell aspect ratio (in the given coordinates, here $\xi$).

The adjacency Gaussian curvature $K_a$ encodes the local disclination density relative to a triangular lattice, i.e., the mesoscopic average of $2\pi(6 - N_i)/6$, where $N_i$ is the number of neighbors of cell $i$ in the tiling\footnote{In an equilateral triangulation, a coordination number $\neq 6$ implies a nonzero angle deficit around a vertex, and hence non-zero Gaussian curvature concentrated at such vertices.} (Fig.~\ref{fig:T1}). 
Note that a flat adjacency metric does not necessarily correspond to a regular hexagonal packing: $K_a$ measures only the \textit{net} density of defects and does not distinguish a regular lattice from a disordered tiling with equal densities of 5- and 7-sided cells. 

We emphasize that the tension and adjacency metrics $\mathbf{g}$ and $\mathbf{a}$ are \textit{distinct} degrees of freedom. Overall, we describe the tissue as a 2D manifold with material coordinates $\bm{\xi}$, on which three metrics encode different pieces of physical information (Fig.~\ref{fig:maps}a):
\begin{enumerate}
    \item The \emph{induced metric} $h_{\alpha\beta}$
    measures the physical distance between  material elements at $\bm{\xi}$ and $\bm{\xi} + d\bm{\xi}$.
    \item The \emph{tension metric} $g_{\alpha\beta}$ measures the active tension across $d\xivec$.
    \item The \emph{adjacency metric} $a_{\alpha\beta}$ measures the number of ``microscopic'' cells along $d\bm{\xi}$, i.e., cell density and aspect ratio.
\end{enumerate}

The discrepancy between these metrics encodes important information (Fig.~\ref{fig:maps}a).
Intuitively, the mismatch between $\mathbf{h}$ and $\mathbf{a}$ measures the cell shape (size and anisotropy) since $\mathbf{h} \propto \mathbf{a}$ if cells (centroids) are equidistant from all of their neighbors (a mathematically precise measure will be introduced below). 
Similarly, the mismatch between
$\mathbf{g}$ and $\mathbf{a}$ measures the local tension magnitude and anisotropy. Microscopically, $\mathbf{g}\propto \mathbf{a}$ exactly when all tensions are equal; $\mathbf{a}$ is made from equilateral ``adjacency'' triangles, $\mathbf{g}$ from potentially anisotropic tension triangles. 
Finally, the stress-metric relationship shows that the strain between the induced metric $\mathbf{h}$ and the tension metric $\mathbf{g}$ determines the macroscopic stress.
Like $\mathbf{a}$ itself, the discrepancies are coarse-grained quantities and measure locally averaged anisotropies.

Together, $\mathbf{a}$ and $\mathbf{g}$ provide the relevant tensorial degrees of freedom for a continuum description of the cell tiling and its mechanics. Both of them are necessary: The ability to control tensions ($\mathbf{g}$) and topology ($\mathbf{a}$) separately distinguishes the tissue from an ordinary fluid foam, where $\mathbf{g} \propto \mathbf{a}$ (all tensions are equal).

\subsubsection{Cell shape and tension anisotropy as Beltrami coefficients}\label{sec:anisotropies_beltrami}

The adjacency metric provides the means to endow local Lagrangian labels $\xi, {\bar \xi}$ with physical meaning. We use $\mathbf{a}$ to introduce a ``topologically informed'' parametrization of the material manifold, namely isothermal coordinates for $\mathbf{a}$.
The cell-shape and the tension anisotropy can then be read off from suitable Beltrami coefficients. Cell rearrangements, which modify $\mathbf{a}$, will lead to a QC flow of the topological parametrization.
In the discrete setting, the $\zeta$-coordinates are closely related to the Koebe-Andreev-Thurston circle packing theorem (see Discussion).

We denote the isothermal coordinates for the adjacency metric $\mathbf{a}$ by $\zeta(\xi,\bar \xi)$, analogous to the $\mathbf{g}$-isothermal coordinates $z$:
\begin{align}
    \bar\partial_\xi \zeta = \mu_a \partial_\xi \zeta \; \Rightarrow a_{\zeta\zeta}=0, \ \ a_{\zeta\bar\zeta} = 2\lambda_a^2
\end{align}
The common Lagrangian cell labels $\xi$ establish a one-to-one correspondence between the two isothermal coordinates $z,\;\zeta$ for the tension and adjacency metrics. 

Since $\mathbf{a}$ and $\mathbf{g}$ are independent metrics, $\zeta(\xi,\bar\xi)$ and  $z(\xi,\bar\xi)$ are independent maps.
Via the chain rule, we switch from the $\xi$- to the $\zeta$-basis: 
\begin{align}\label{eq:mapping_summary_2}
    \xi \xmapsto{\mu_{\xi|\zeta}} \zeta(\xi,\bar\xi) \xmapsto{\mu_{\zeta|z}} z(\zeta, \bar \zeta) \xmapsto{\mu_{z|w}} w(z,\bar z)
\end{align}

The relationships between the embeddings and maps between them are graphically represented in Fig.~\ref{fig:maps}b. The first map  $\xi\mapsto\zeta$ goes to the parametrization $\zeta$ in which $\mathbf{a}$ is isotropic ($\mathbf{a}$-isothermal coordinates). The second map $\zeta\mapsto z$ makes $\mathbf{g}$ isotropic, defining the stress-free reference state. The final map $z\mapsto w$ to the physical configuration is determined by force balance, $\partial_\alpha\sigma_{\alpha\beta} = \partial_\beta p$.
The Beltrami coefficient $\mu_{\zeta |z}$ is defined by the isothermal embedding of $\mathbf{g}$, parametrized in $\zeta$ coordinates.

The adjacency coordinates $\zeta$ serve as a convenient set of coordinates adapted to the adjacency of the cell network -- in $\zeta$-coordinates, cells have isotropic shapes and density $n_0 \lambda_a^2$. 
If interfacial tensions are perfectly isotropic, then $\mathbf{a}\propto\mathbf{g}$ and $z \propto \zeta$. More generally, the Jacobian $\partial_{\boldsymbol{\zeta}} z$ of the QC map from $\zeta$ to the tension reference state $z$ defines the local tension configuration~\cite{Brauns.etal2024,Claussen.etal2024}.
Microscopically, $\partial_{\boldsymbol{\zeta}} z$ deforms equilateral ``adjacency'' triangles into anisotropic tension triangles.
In particular, the Beltrami coefficient $\mu_{\zeta |z} = \bar\partial_\zeta z / \partial_\zeta z=\mu_g(\zeta,\bar{\zeta})$ encodes the magnitude and direction of the tension anisotropy.
Overall, in $z$-coordinates, the adjacency metric specifies the local cell density and the tension anisotropy.

By contrast, the map from $z$ to the physical configuration $w$, via its Beltrami coefficient $\mu_{z|w} = \bar\partial_z w / \partial_z w$, determines the elastic shear and macroscopic deviatoric stress. In fact, $\mu_{w|z}$ is proportional to deviatoric stress [cf.\ Eq.~\eqref{eq:stress_w}].
The total anisotropy of the map $w(\zeta,\bar\zeta)$ from isotropic adjacency triangles to the physical configuration is given by the chain rule:
\begin{align}
    \mu_{\zeta |w} &= \frac{\bar \partial_\zeta w}{\partial_\zeta w} = \frac{
    \bar\partial_\zeta z \partial_z  w + \bar \partial_\zeta \bar z \bar\partial_z w 
    }{\partial_\zeta z \partial_z w + \partial_\zeta \bar z \bar \partial_z w } \nonumber \\
    &= \frac{\mu_{\zeta|z} + e^{-i2\phi_{\zeta|z}} \mu_{z|w} }{1+  e^{-i2\phi_{\zeta|z}} \bar\mu_{\zeta|z} \mu_{z|w}} \\
    &\approx \mu_{\zeta |z} + e^{-i2\phi_{\zeta|z} } \mu_{z|w}
    \label{eq:Beltrami_composition}
\end{align}
where $e^{-i2\phi_{\zeta|z}} = \overline{\partial_\zeta z}/ \partial_\zeta z$, and we assumed $|\mu_{\zeta|z}|,|\mu_{z|w}|\ll 1$ in the last step.

Taken together, the $\mathbf{a}$-isothermal coordinates $\zeta$ allow us to define the local tension configuration in the continuum as the Jacobian of the map $z(\zeta)$. Per Sec.~\ref{sec:statics}, the macroscopic stress is \emph{independent} of the adjacency metric and hence of tension anisotropy $\mu_{\zeta|z}$. However, the latter plays an important role in cell rearrangement, as we will see below.

\subsubsection{Kinematics of cell rearrangement: Flip flow}\label{sec:kinematics}

Dynamics of the adjacency metric are driven by the (oriented) T1 rate $R(\xi, \bar{\xi}, t)$, to which individual T1s make infinitesimal contributions. 
These increments are most easily expressed in Eulerian $\zeta(t)$-coordinates
\begin{align}
    \mathbf{a}\bigl(\bm{\zeta}(t), t\bigr) \mapsto  \mathbf{a}\bigl(\bm{\zeta}(t), t\bigr) + \partial_t \mathbf{a}\bigl(\bm{\zeta}(t), t\bigr) dt
\end{align}
As above, we write $\zeta=\zeta(t)$ for short in the following.
Since T1s do not change the cell number, they correspond to pure shear deformation of $\mathbf{a}$:
\begin{align}
    \label{eq:T1_rate_definition}
    \partial_t \lambda_a(\zeta,\bar\zeta, t)= 0, \quad  \partial_t \mu_a(\zeta,\bar\zeta, t) := R(\zeta,\bar\zeta, t)
\end{align}
In the present framework, Eq.~\eqref{eq:T1_rate_definition} is the definition of the rate of topological rearrangement in the continuum.
We introduced $R$ as a complex Beltrami coefficient; it can also be represented by a (trace-free) Cartesian tensor $\mathbf{R}$ analogous to $\mathbf{m}$ in Eq.~\eqref{eq:m_definition} with $R_{zz} = 4R$ and $R_{z\bar{z}} = 0$.

\begin{figure}
    \centering
    \includegraphics[scale=0.75]{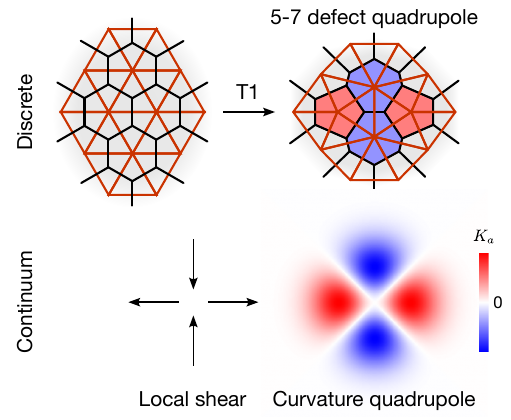}
    \caption{In a hexagonal lattice, a T1 transition generates two 5-7 defect pairs. Correspondingly, a localized shear generates a curvature quadrupole in the adjacency metric $\mathbf{a}$.}
    \label{fig:T1}
\end{figure}

For example, consider a localized ``rearrangement event'' $d a_{\zeta\zeta} = 4\lambda_a^2 R \, dt$, $R = c \ell_R^2 e^{-\zeta\bar \zeta/(2\ell_R^2)} $ with spatial extent $\ell_R$, on an initially uniform background $\lambda_a=1$.
The complex coefficient $c$ sets the shear orientation and magnitude.
To mimic an ``isolated T1'' in a hexagonal lattice, one sets $\ell_R$ to the scale of the cell size, and $|c|\sim 1/t_\mathrm{T1}$, where $t_\mathrm{T1}$ is the timescale of the T1 process.
Microscopically, such a T1 creates two 5-7 defect pairs starting from a background of six-sided cells (Fig.~\ref{fig:T1}, top).
In the continuum, defects are localized quadrupoles in the adjacency curvature $K_a$.
Indeed, using the St.~Venant formula Eq.~\eqref{eq:St-Venant}, the increment $da_{\zeta\zeta}$ leads to:
\begin{align}
    d K_a = \ell_R^{-2} \Re\!\left[ c \bar \zeta^2 \right] e^{-\zeta\bar \zeta/(2\ell_R^2)} \, dt
\end{align}
For $\arg c=0$, $dK_a\propto(\zeta_1^2-\zeta_2^2)  \, dt$ (Fig.~\ref{fig:T1}, bottom).
Note that upon coarse-graining, local curvature quadrupoles average out.
Information about individual cell coordination numbers is lost, and only the local disclination density is retained, i.e.\ the deviation of the \emph{average} coordination number from 6.

The T1-induced shear deformation Eq.~\eqref{eq:T1_rate_definition} requires an update to the isothermal $\zeta$-coordinates, $\zeta(t) \mapsto \zeta(t+dt) = \zeta(t) + \dot{\zeta}(\zeta(t), \bar\zeta(t), t) dt$. We refer to $\dot\zeta(t)$ as the \emph{flip flow}, since it corresponds to edge flips in the discrete triangulation. The flip flow is QC and obeys:
\begin{align}
   \bar\partial_\zeta {\dot{\zeta}}(\zeta,\bar\zeta, t)  = \partial_t \mu_a (\zeta,\bar\zeta, t) = R(\zeta,\bar\zeta, t)
    \label{eq:Beltrami_flow_a}
\end{align}

We now compute how the flip flow changes the local tension anisotropy.
The time derivative of $\mathbf{g}(\bm{\zeta})$ combines intrinsic active tension dynamics $\mathbf{m}$ and cell rearrangement $R$.
The former leads to reference state flow $z\mapsto z +\dot{z}\,dt$, and the latter to the coordinate transformation $\zeta\mapsto \zeta +\dot{\zeta}\,dt$. The flip flow thus enters as a continuous ``reparametrization'' of the tension manifold.
The dynamics of the Beltrami-coefficient $\mu_{\zeta|z}$, assuming $|\mu_{\zeta|z}|\ll 1$, are given by:
\begin{align}\label{eq:advection_linearized}
    \partial_t \mu_{\zeta | z} + &2\left(\mathrm{Re}[\dot{\zeta} \partial_\zeta]\mu_{\zeta|z}- i \mu_{\zeta | z} \mathrm{Im}[\partial_\zeta {\dot{\zeta}}] \right) \approx \nonumber \\
    &\quad \approx \partial_t \mu_g(\zeta, \bar\zeta) - \partial_t \mu_a(\zeta, \bar\zeta)
\end{align}
The bracketed terms on the LHS represent advection and co-rotation and will be suppressed in the following.
The RHS arises from the QC transformations of $z$ and $\zeta$ (see Appendix~\ref{app:beltrami-kinematics_1} for the derivation; Fig.~\ref{fig:Beltrami-flip} illustrates the underlying geometry).

In summary, T1s change the adjacency metric $\mathbf{a}$, which we represent via the flip flow $\dot \zeta$.
The flip flow is a reparametrization that changes the tension anisotropy, but has no \emph{instantaneous} effect on the tension metric $\mathbf{g}(\xivec)$  in fixed, Lagrangian coordinates.
In our framework, cell rearrangement is thus a purely kinematic process. 
This is best motivated by looking at physical space.
Microscopically, a T1 flips an edge whose physical length $\ell_{ij}=0$, and is hence ``invisible'': it has no instantaneous effect on physical shape or macroscopic stress.
We next describe a specific model for the \emph{dynamic} change of the tension manifold elicited by the \emph{kinematic} effect of T1s on tension anisotropy.

\subsubsection{Dynamics of cell rearrangement: Active and passive T1s}\label{sec:T1_threshold}

To fill the kinematic relations \eqref{eq:Beltrami_flow_a}--\eqref{eq:advection_linearized} with life, we need a continuum law for the rate and orientation of T1s, and a prescription for the tension dynamics caused by rearrangements.

Phenomenologically, T1s relax the overall cell shape anisotropy by allowing cells to rearrange~\cite{Blanchard.etal2009}. In the framework of Sec.~\ref{sec:three_metrics}, cell shape anisotropy ($\mathbf{a}$ to $\mathbf{h}$ discrepancy) can be decomposed into tension anisotropy ($\mathbf{a}$ to $\mathbf{g}$ discrepancy) and emergent elastic stress ($\mathbf{g}$ to $\mathbf{h}$ discrepancy, i.e., strain relative to the stress-free reference). These distinct contributions are not accounted for in previous phenomenological models~\cite{Popovic.etal2017,Duclut.etal2022}.

We will argue that T1s can be driven by local tension dynamics $\mathbf{m}$ (``active T1s'') or by strain due to external forces (``passive T1s''). T1s then relax cell-shape anisotropy, to which both stress and tension anisotropy contribute.
However, we do \emph{not} predict that T1s drive cells towards a regular packing, or generically remove topological defects. Continued anisotropy generation from active tension dynamics and a non-zero yield threshold for T1s mean that cell-shape anisotropy can remain non-zero despite relaxation.
Indeed, in past work, we used simulations and mean-field theory to show that active T1s can drive proliferation of defects, causing a regular packing to become disordered~\cite{Claussen.etal2024,Claussen.Brauns2025}.
We found a steady state where anisotropy drive and T1 relaxation balance. If T1s are oriented by a suitable cue, this gives rise to a steady shear flow at the continuum scale.

As a minimal model for topological plasticity, we thus propose to represent T1s by continuous relaxation of the total anisotropy $\mu_{\zeta|w}$.
Eq.~\eqref{eq:Beltrami_composition} shows that $\mu_{\zeta|w}$ combines the tension and stress anisotropies.
The (oriented) rearrangement rate, therefore, reads
\begin{align}\label{eq:T1_rate}  
    R(\zeta,\bar\zeta)  &= -\gamma_{\mathrm{T1}} \frac{\delta |\mu_{\zeta|w}|^2}{\delta \bar{\mu}_a} =  \gamma_{\mathrm{T1}}\big(\mu_{\zeta|z}, \mu_{z|w},...\big) \mu_{\zeta|w} 
\end{align}
Here, $\gamma_{\mathrm{T1}}$ is a rate coefficient (the ``T1 rate''), which depends on the local distribution of microscopic tension configurations, as we will discuss below.
To leading order, this can be summarized by a dependence on the mean tension and stress anisotropy.

The variation of $|\mu_{\zeta|w}|^2$ follows from the Beltrami composition formula~\eqref{eq:Beltrami_composition} (see Appendix~\ref{app:beltrami-kinematics_2}).
The total anisotropy decomposes into the tension anisotropy and the stress anisotropy
\begin{equation}
    \mu_{\zeta|w} \approx \mu_{\zeta|z} + e^{-i2\phi_{\zeta|z}}\mu_{z|w}.
\end{equation}
T1s can thus be driven by internal tension changes or by elastic deformations due to external forces~\cite{Brauns.etal2024,Claussen.etal2024}.
Accordingly, we define active and passive T1 rates
\begin{subequations}
\begin{align}
    R_\mathrm{act} &= \gamma_\mathrm{T1} \mu_{\zeta|z} \\
    R_\mathrm{pass} &= \gamma_\mathrm{T1} \mu_{z|w} \label{eq:T1_rate_pass}
\end{align}
\end{subequations}
so that $R = R_\mathrm{act} + e^{-i2\phi_{\zeta|z}} R_\mathrm{pass}$.
Passive T1s are driven by anisotropic elastic strain $\mu_{z|w}$ (and hence stress anisotropy $\sigma_{ww}$). By contrast, active T1s can occur in the absence of macroscopic stress. Instead, they are driven by the tension anisotropy $\mu_{\zeta|z}$.

Thus, Eq.~\eqref{eq:T1_rate} qualitatively reproduces the mechanisms driving T1s on the cell level~\cite{Brauns.etal2024,Claussen.etal2026a}. There, the combined tension (active) and strain (passive) anisotropy determines when a cell-cell interface reaches zero length $\ell_{ij}=0$, triggering a T1.
Ref.~\cite{Claussen.etal2026a} worked out the critical \emph{microscopic} Beltrami coefficients for which $\ell_{ij}=0$ geometrically\footnote{Note that the quantity in the microscopic threshold Eq.~\eqref{eq:microscopic_T1} differs from the total Beltrami coefficient that appears in the phenomenological Eq.~\eqref{eq:T1_rate}. Indeed, $2\mu_{\zeta|z}^\triangle+e^{-2i\phi_{\zeta|z}}\mu_{z|w}^\triangle \neq  \mu_{\zeta| w}^\triangle \approx \mu_{\zeta|z}^\triangle+e^{-2i\phi_{\zeta|z}}\mu_{z|w}^\triangle$ by Eq.~\eqref{eq:Beltrami_composition}. On a microscopic level, ``active'' tension anisotropy contributes twice as much to T1s as ``passive'' cell shape strain.
This fact can be accommodated in our framework, which decomposes the total cell shape anisotropy into tension- and strain-anisotropy.}. To linear order:

\begin{align}
    \label{eq:microscopic_T1}
    \left\vert 2\mu_{\zeta|z}^\triangle+e^{-2i\phi_{\zeta|z}}\mu_{z|w}^\triangle \right\vert = \tfrac{1}{2}
\end{align}
at the T1 threshold.
Here, the superscript $\triangle$ marks quantities on the single-triangle level. The Beltrami coefficients $\mu_{\zeta|z}$ and $\mu_{z|w}$ entering the continuum theory represent a macroscopic average over many tension triangles. The microscopic T1 threshold Eq.~\eqref{eq:microscopic_T1} implies the existence of a yield strain for passive T1s: below a critical $|\mu_{z|w}|$, the T1 rate $\gamma_{\mathrm{T1}}$ vanishes.
This implies, in particular, that a finite level of anisotropy can be stable if it is too small to trigger T1s.
More generally, $\gamma_{\mathrm{T1}}$ depends on the microscopic distribution of cell-level configurations $\mu_{\zeta |w}^\triangle$, which must be summarized by a suitable set of local order parameters. A mean-field approach~\cite{Claussen.Brauns2025} allows systematic calculation of $\gamma_{\mathrm{T1}}$ in terms of these order parameters, together with equations governing their dynamics.

\subsubsection{T1s relax tension- and stress anisotropy}\label{sec:T1_relax}

To complete the model for plasticity in the continuum, we need to address how cell rearrangements impact the tension metric $\mathbf{g}$. \emph{A priori}, the flip flow due to cell rearrangements is a reparametrization: the tension manifold remains invariant (the macroscopic stress, depending only on $\mathbf{g}$ and the physical BCs, also remains invariant). Edge flips in the triangulation pass between different triangulations that belong to the same stress state.
However, T1s are not only an abstract ``remeshing'' of the tension manifold; they create a new cell interface whose tension dynamics are governed by specific (bio-)physical processes at the microscopic level.
The model below for the effect of T1s on the tension metric assumes that the new interface's tension takes on an isotropic value. This is exactly true in a conventional fluid foam where surface tensions are always isotropic, and is a reasonable null model for a tissue.
More precisely, by Eq.~\eqref{eq:advection_linearized}, edge flips change the tension anisotropy $\mu_{\zeta|z}$ (this is geometrically evident in the discrete setting; Fig.~\ref{fig:Beltrami-flip}). The T1-induced change of tension anisotropy can then elicit subsequent dynamics of $\mathbf{g}$ through the processes that govern the microscopic tensions.

For simplicity, this section linearizes for small strain and tension anisotropy, $|\mu_{\zeta|z}|,|\mu_{z|w}|\ll 1$, and further $\lambda_g\approx 1$. We can thus neglect the distinctions between the different coordinate systems $\zeta, z, w$, and express everything in $z$-coordinates.
As a minimal model, we propose that passive T1-transitions remodel the tension metric, in addition to the active ``drive'' provided by $\mathbf{m}$.
\begin{align}\label{eq:T1_g_change}
    \partial_t \log\mathbf{g} = \mathbf{m} + \mathbf{R}_\mathrm{pass}
\end{align}
Hence, for the Beltrami coefficient (the tension anisotropy):
\begin{align}
    \partial_t \mu_g = \frac{m_{zz}}{4}+ R_\mathrm{pass} = \frac{m_{zz}}{4}+ \gamma_\mathrm{T1} \mu_{z|w}
\end{align}
Here, only $\mathbf{R}_\mathrm{pass}$ and not $\mathbf{R}_\mathrm{act}$ contributes; the reason for this will become clear in the remainder of this section.
While T1s reduce anisotropy, $\mathbf{m}$ can subsequently build it up and drive further T1s.

As we now show, Eq.~\eqref{eq:T1_g_change} captures the physically expected behavior for both active and passive plastic flow.
Using the passive T1-rate Eq.~\eqref{eq:T1_rate_pass}, and relating deviatoric strain $\mu_{z|w}$ to the stress via Eq.~\eqref{eq:stress_w}: 
\begin{align}
    \label{eq:T1_g_change_stress}
    \partial_t \mu_g %= \frac{m_{zz}}{4} +  \gamma_\mathrm{T1} \mu_{z|w} 
    =  \frac{m_{zz}}{4} + \gamma_\mathrm{T1} \frac{\sigma_{zz}}{2p_0 \lambda_g}
\end{align}
By a slight abuse of notation, we write $\sigma_{zz}$ for the elastic stress in $z$-coordinates, which, of course, depends on the deformation $w$.
On the other hand, changing the tension metric at fixed physical configurations changes the stress.
We evaluate the Cauchy stress Eq.~\eqref{eq:stress-metric-relation} in the current configuration $z$, but using the time-incremented tension metric $\mathbf{g}(t+dt)$.
An anisotropy increment $\delta \mu_g$ shifts the reference configuration, so $\mu_{z'|w} = \mu_{z|w} + \mu_{z'|z} =  \mu_{z|w} - \delta \mu_g$. By Eq.~\eqref{eq:stress_w}, the resulting stress increment is $\delta \sigma_{zz} = 2p_0 \delta \mu_{z'|w}$, so (to linear order):
\begin{align}\label{eq:stress-expansion-Eulerian_z}
     \partial_t \sigma_{z z}\big|_{w = \mathrm{const.}} = - 2p_0\lambda_g \,\partial_t \mu_g
\end{align}
Together, Eqs.~\eqref{eq:T1_g_change_stress},~\eqref{eq:stress-expansion-Eulerian_z} result in effective Maxwell relaxation, with a source provided by $\mathbf{m}$:
\begin{align}\label{eq:maxwell_relax}
    \partial_t \sigma_{zz}\big|_{w = \mathrm{const.}} = -\gamma_\mathrm{T1} \sigma_{zz} -\tfrac{1}{2}p_0 m_{zz}
\end{align}
Eq.~\eqref{eq:maxwell_relax} describes how passive T1s relax stress when the cell positions $w$ are pinned. Note that on the microscopic level, this corresponds to pinning cell centroids $\mathbf{r}_i$; cell vertices $\mathbf{r}_{ijk}$ are still free to move to establish local force balance. While T1s relax only the anisotropic stress, due to mechanical balance Eq.~\eqref{eq:balance_w}, vanishing anisotropic stress also implies uniform isotropic stress.

For the reference-state velocity field, the combination of stress relaxation and active driving manifests as an effective Stokes flow [cf.\ Eq.~\eqref{eq:effective_Stokes}]:
\begin{align}
    \Delta \dot{\mathbf{z}} &= \frac{2\gamma_\mathrm{T1}}{p_0}
    \nabla\cdot \dev\bm{\sigma} + \nabla\cdot \dev\mathbf{m}
 \label{eq:effective_stokes_with_T1s}
\end{align}
The anisotropic macroscopic stress enters with an effective viscosity $p_0/(2\gamma_\mathrm{T1})$. Note that in many cases, $\gamma_\mathrm{T1}$ may vanish below a critical yield strain~\cite{Brauns.etal2024}.
Eq.~\eqref{eq:effective_stokes_with_T1s} thus combines active flow due to tension dynamics $\mathbf{m}$ with viscous relaxation of elastic stress due to passive T1s.
The limiting cases discussed so far (free boundaries with flat tension metric; pinned cells) are particularly simple since one does not need to solve the effective elastic problem Eq.~\eqref{eq:flow_w} for $\dot{w}_\beta$ (Sec.~\ref{sec:adiabatic_w}). In general, one must solve this non-local problem, as formally stated in Eq.~\eqref{eq:functional_derivative}, find the physical configuration $w$ and thus the T1 rate $R = \gamma_\mathrm{T1} \mu_{\zeta|w}$.

We now explore the physical consequences of the proposed model in the purely active and passive cases.
Consider a conventional fluid foam, where $\mathbf{m}=0$, as the paradigmatic case for passive T1s. 
External forces can build up deviatoric strain and thus macroscopic stress $\bm{\sigma}$, while microscopic tensions remain constant. Eventually, strain causes T1s that relax macroscopic stress.
Eq.~\eqref{eq:effective_stokes_with_T1s} reproduces this physical intuition. 
Further, since all tensions equal the fluid's surface tension, $\mathbf{g} = \mathbf{a}$: T1s should directly modify the tension metric. This is automatically ensured by the dynamics Eq.~\eqref{eq:T1_g_change}: If initially $\mathbf g\propto\mathbf a$, then $\mu_{\zeta|z}=0$ and $R_{\rm act}=0$, so $\partial_t\mu_g=R_{\rm pass}=R=\partial_t\mu_a$, and $\mathbf g\propto\mathbf a$ is preserved.

We next consider the purely active case, $w = z$, i.e.\ $\mu_{z|w} = 0$. Here, T1s happen when $|\mu_{\zeta|z}|$ increases due to driving by $m_{z z}$ (for example, positive mechanical feedback \cite{Brauns.etal2024,Claussen.etal2024} or a morphogen gradient \cite{Ibrahimi.Merkel2023,Claussen.Brauns2025}).
Substituting Eq.~\eqref{eq:T1_g_change} into Eq.~\eqref{eq:advection_linearized} shows that T1s relax tension anisotropy: 
\begin{align}
    \frac{d}{dt} \mu_{\zeta|z} = \tfrac14 m_{zz} -R_\mathrm{act}
\end{align}
Active T1s thus limit the tension anisotropy a tissue can attain.
Steady active plastic flow results when(a persistently oriented) driving $\mathbf{m}$  and relaxation balance at a finite T1 rate.
Active flow can take place at vanishing macroscopic stress and is therefore fundamentally different from conventional Maxwell viscoelasticity.
Instead, it is characterized by sustained microscopic tension anisotropy, experimentally detectable as anisotropic junctional myosin~\cite{Streichan.etal2018}.

To summarize, the present framework describes a tissue by two independently controlled metrics $\mathbf{a},\,\mathbf{g}$. A kinematical prescription determines what happens during topological change (dynamics of $\mathbf{a}(t)$, keeping $\mathbf{g}(t)$ fixed). The dynamics of the tension metric $\mathbf{g}$ depend on the specific model or biological context of interest.
We showed that a minimal and microscopically motivated choice, Eq.~\eqref{eq:T1_g_change}, naturally gives rise to Maxwell-like stress relaxation through passive T1s while active T1s relax tension anisotropy.

\section*{Discussion}
\setcounter{subsubsection}{0}

We have presented a continuum theory for active tension nets, a paradigmatic example of an \emph{active solid}, where the local stress, rather than the local target shape, is the physical (or biological) control parameter. 
Our key finding is that internal, active forces alone, without the need for a passive elastic ``background'', define a physical shape that elastically resists deformations. 
This explains how tissues can maintain their shapes even when passive stresses rapidly dissipate through molecular turnover.
Our theory, therefore, describes a class of \textit{active solids} not based on conventional continuum mechanics, in contrast to previous works where active stresses are added on top of a conventional (passive) elastic background~\cite{Maitra.Ramaswamy2019,Baconnier.etal2022,Brauns.etal2026}.

Our theory shows how cell-level tensions determine macroscopic stress and shape.
Building on these results, we explain how tissues can ``morph'' via adiabatic dynamics of cell-level tensions while retaining rigidity against external forces. This is in stark contrast to the idea that tissue flows require  ``fluidization''~\cite{Bi.etal2015,Tetley.etal2019,Wang.etal2020}. 
The capacity to change shape while retaining shear rigidity is likely advantageous for living matter (e.g., for robustness), while local, subcellular tension may be more easily sensed and controlled than cell shape and macroscopic stress.

\subsubsection{Riemannian geometry of an active solid}

To represent the microscopic configuration of active tensions in the continuum, we introduced a Riemannian \emph{tension metric} $\mathbf{g}$.
The notion of a tension metric to represent the microscopic active force configuration is one key technical innovation of this work.
The tension metric, together with the mechanical BCs, determines the macroscopic stress tensor and the positions adopted by cells in mechanical equilibrium.
By identifying the tension metric as the relevant macroscopic degree of freedom of active tension nets, our theory explains how local contractile activity translates into large-scale shape.

Crucially, many microscopic tension triangulations specify the \emph{same} tension metric. They correspond to reparametrizations of the tension manifold and do not affect physical stress.
Put differently, the macroscopic (i.e., tissue-scale) mechanical state only depends on the shape of the tension manifold, not how it is discretized by ``painting'' individual tension triangles onto it, much like in a finite-element discretization of a partial differential equation.
In particular, different triangulations of a region in the plane all define the same tension metric -- only the shape of the region's boundary is relevant.
Concretely, the macroscopic stress is \emph{independent} of the local tension anisotropy, and instead reflects external forces and ``incompatibility'' between cell and tension density. This can be tested in experiments, where macroscopic stress can be inferred from tissue-scale laser ablation~\cite{Saadaoui.etal2020}.

In contrast to the tension metric and macroscopic stress, microscopic cell shapes and tension anisotropy depend on how the tension manifold is triangulated, i.e., the local cell network topology. Morphogenesis is ultimately controlled at the cellular level, so cell shape and local tension configuration are important variables.
This raises the challenge of incorporating discrete topology into a continuum theory: a continuum model requires information in addition to $\mathbf{g}$. 
The most important piece is the \emph{adjacency metric} $\mathbf{a}$, which measures the local, ``oriented cell density'' -- the number of cells traversed along an infinitesimal displacement vector.
Its Gaussian curvature is determined by the local disclination density of the cell packing.
The dynamics of the adjacency metric describe cell rearrangements.
While the tension metric $\mathbf{g}$ determines the macroscopic mechanics, the local tension anisotropy is encoded in the ``mismatch'' of $\mathbf{g}$ and $\mathbf{a}$ -- the shape of tension triangles.
To mathematically describe rearrangement and tension anisotropy in a principled way, we introduce the isothermal embedding $\zeta(\xi,\bar\xi)$ of the adjacency metric $\mathbf{a}$, which serves as a ``topologically informed'' reference parametrization.
The local tension configuration (magnitude and anisotropy) is then given by the Jacobian $\partial_{\bm{\zeta}}\mathbf{z}$.

In this framework, tissue dynamics correspond to changing tension and adjacency metrics.
We describe tissue flow driven by adiabatic dynamics of active forces by tracking the reference state $z(\xi,\bar\xi, t)$, defined through isothermal coordinates of the evolving tension metric $\mathbf{g}(t)$.
The resulting QC \emph{adiabatic flow} $\dot{z}$ is governed by Eq.~\eqref{eq:beltrami_flow_Eulerian}. Large, plastic deformations entail topological cell rearrangement, i.e., changes in $\mathbf{a}(t)$. We similarly track the evolving ``topological'' reference parametrization $\zeta(\xi,\bar\xi, t)$, and cell rearrangement becomes the \emph{flip flow} $\dot{\zeta}$. Tension dynamics, plasticity, and elastic deformation can thus be formulated as two coupled QC flows, Eqs.~\eqref{eq:beltrami_flow_Eulerian} and~\eqref{eq:Beltrami_flow_a}.

Overall, the present theory comprises the minimal description of tension nets and \emph{active plasticity} in the continuum: The tension metric $\mathbf{g}(\xivec)$ 
and the adjacency metric $\mathbf{a}(\xivec)$.
The physical positions $\rvec(\xivec)$ of material elements define the induced metric and are determined by force balance.
The discrepancies between the three metrics $\mathbf{a}, \, \mathbf{g}, \; \mathbf{h}$ systematically quantify the anisotropies of microscopic shape, active forces, and macroscopic stress (Fig.~\ref{fig:maps}).
We believe that the systematic description of stress and network topology in terms of Riemannian geometry and QC maps will prove useful to study other materials with dynamic microscopic structure, such as granular matter~\cite{Lazar.etal2012} (more on this below).

\subsubsection{Emergent metric elasticity}

On a formal level, the present theory has parallels with metric elasticity, where a Riemannian \emph{target metric} represents the rest lengths between adjacent material points~\cite{Efrati.etal2009,Efrati.etal2013,Kupferman.etal2015}. Deviations from these rest lengths in the physical configuration cost elastic energy. More precisely, the mismatch of the target metric and the induced metric of the physical configuration defines a strain from which stress follows through a prescribed constitutive relation (or, equivalently, an elastic energy functional).

Like metric elasticity, the present theory represents the intrinsic state of the tissue using Riemannian metrics. However, instead of rest lengths, these metrics represent internal tensions and cell adjacency.
A stress-free reference state and an effective stress-strain relationship then \emph{emerge} from the intrinsic tension configuration. The \emph{stress-metric relation} determines the physical stress tensor from the embedding of the tension metric given by the cell positions, i.e., the arrangement of active force dipoles in physical space. 
In addition to the tension metric, the adjacency metric is key to resolve the relevant microstructure in the continuum. Tension configurations and cell shapes are represented by the discrepancies between adjacency metric and the tension and physical metric, respectively.

In metric elasticity, the mechanical equilibrium can be residually stressed when the curvature of the target metric cannot be realized by the physical metric. The present theory also features residual stresses; however, they do not arise from curvature per se, but when cell- and tension densities are incompatible (to which curvature can contribute). In this case, the equation of state cannot realize the pressure required by the stress-free reference state, and the tissue adopts a force-balanced, but residually stressed configuration.

Overall, the present theory furnishes the continuum limit of cell-level models dominated by \emph{active} mechanics. 
By contrast, conventional metric elasticity is well-suited to describe the continuum limit of passive models (generalized spring networks).
For instance, in the context of the well-known ``area-perimeter'' vertex model~\cite{Farhadifar.etal2007}, Ref.~\cite{Grossman.Joanny2022} defined a network tensor that plays the role of the elastic reference metric. In this setting, plasticity can be modeled by relaxational dynamics of the reference metric \cite{Efrati.etal2013}.

The present theory is therefore not a generalization or variant of metric elasticity, but describes a distinct physical system (an active solid). Nevertheless, the theories share several mathematical features, notably the use of differential geometry.
In this work, we have focused on the mechanics, topology, and flow of active solids within a fixed, flat plane.
In metric elasticity, many interesting phenomena (e.g., wrinkling due to geometric incompatibility) arise when two-dimensional shells are allowed to deform out of plane~\cite{Efrati.etal2009}.
Similar questions may be studied for active solids, as we discuss next.

\paragraph*{Tissues on curved and dynamic surfaces.}

For simplicity, we considered a tissue patch with a disk topology, embedded in the plane.
The geometric nature of the theory allows generalizing to tissues on fixed but curved surfaces, such as the ellipsoidal blastoderm of the \emph{Drosophila} embryo.
We explain the required modifications in App.~\ref{app:curved}.

In brief, the Euclidean metric in Eq.~\eqref{eq:g-isothermal} is replaced by the metric of the curved physical space. Centrally, we explain how to construct the effective reference state via a conformal map from the tension manifold to the physical surface, like in the flat case.
The in-plane force-balance condition $\nabla_\alpha\sigma_{\alpha\beta}^\mathrm{tot}=0$ must also be formulated in a covariant manner~\cite{Salbreux.Julicher2017}, which modifies the linearized elasticity equations of Sec.~\ref{sec:linearized}.
Using a common set of isothermal coordinates for the tension metric and the metric of the physical surface leads to rather simple expressions. For illustration, we consider a tissue patch on a hemisphere.
The curvature of the physical surface acts like a fixed background field, to which tissue mechanics is coupled. This represents a potential mechanism for geometry sensing, e.g., to position tissue relative to regions with specific curvature~\cite{Neipel2024}.

If the tissue topology is non-trivial, new geometric modes and invariants appear. For example, on an annulus, there is a curl-free vector field that cannot be written as a potential gradient~\cite{Lee2012}, and the annulus' ``aspect ratio'' is conformally invariant~\cite{Ahlfors1966}.
Due to the important role of curl-free and conformal modes in our theory, this may have interesting physical consequences to be explored in future work.

In the above setting, the cells are still constrained to lie on a fixed (albeit curved) physical surface. Studying tissues that form dynamic surfaces, free to deform \emph{out of plane}, is an interesting and non-trivial task for future work.
This generalization may work similarly to the membrane theory of shells~\cite{Landau.Lifshitz1986}.
Out-of-plane deformations provide one additional degree of freedom, the normal displacement.
In addition to in-plane force balance, one must account for normal force balance, which provides one additional constraint.
The latter couples in-plane stress and extrinsic physical curvature to normal pressure gradients via the Laplace law in the form $\sigma_\mathrm{tot}^{\alpha\beta} b_{\alpha\beta} = p_\perp$, where $b_{\alpha\beta}$ is the second fundamental form (containing the principal curvatures), and $p_\perp$ is the \emph{bulk} pressure drop across the surface. 
%\fb{I removed the $\Delta$ since we use it for the Laplace operator everywhere else.}
Existence and uniqueness of balanced states are, however, not obvious, and depend on the physics of the situation (boundary conditions, additional bending resistance, incompressibility of volume enclosed by the tissue sheet).

We hope that future work can build on the ideas presented here to carry out this interesting generalization. It is required to describe situations where in-plane contractility drives three-dimensional shape change of tissue sheets, e.g., during visceral organ morphogenesis~\cite{Mitchell.etal2022b}.

\subsubsection{Implications for active-matter models of morphogenesis}

Taking a step back, the present theory is a marked departure from an existing class of continuum models of active matter and tissue mechanics, notably regarding the nature and role of ``active stress''.
``Active fluid'' models posit a phenomenological active stress tensor~\cite{Streichan.etal2018,Saadaoui.etal2020,Serra.etal2023}, assumed to be proportional to the local motor molecule concentration. Flow is driven by the net active forces, balanced against the passive stress of a viscous ``background'' with vanishing shear modulus.
The stress anisotropy is proportional to the local tension anisotropy~\cite{Streichan.etal2018,Serra.etal2023}, and the flow speed is set by the balance of motor-generated force and viscosity.

By contrast, in the present model, plastic flow is caused by dynamic changes of tensions. Flow speed is set by the \emph{rate of change} of motor molecule concentration, and \emph{independent} of the overall strength of the active force. This prediction can be tested in future experiments, for example by changing the (effective) tissue viscosity or slowing down motor-molecule turnover.
Moreover, in the present model flow is adiabatic: the stress remains balanced throughout.
The system remodels from one solid configuration to the next and thus ``flows while remaining solid''\footnote{Note, however, that the yield strain becomes highly anisotropic during active plastic flow. Due to the ongoing T1 transitions, there is a large number of cells near the T1 threshold, and the yield strain parallel to tension anisotropy can become very small.}. 
In particular, the macroscopic stress remains isotropic during tissue flow (in the absence of external forces), even if tensions are anisotropic, unlike in active fluid models. 
This could be tested by laser ablations on supracellular scales~\cite{Saadaoui.etal2020}.

A steady-state tension anisotropy, distinct from stress anisotropy, can emerge from the balance of anisotropy-increasing processes (such as mechanical feedback or morphogens) and resetting by T1s. These predictions rely on the separation of timescales between morphogenetic flow and relaxation of passive stress and mechanical equilibration. It will be important to explore what happens if this assumption is relaxed.

On a more conceptual level, in the present model, there is no passive shear stress (all viscous and passive contributions are assumed to relax rapidly).
Local motor molecule concentration is encoded in the tension metric $\mathbf{g}$ instead of an active stress tensor that is conventionally invoked in active matter theories \cite{Marchetti.etal2013,Maitra.Ramaswamy2019,Brauns.etal2026}. Indeed, the tensile stress $\bm{\sigma}$ depends both on $\mathbf{g}$ and the embedding $\mathbf{r}(\xivec)$ of active force dipoles into physical space via the stress-metric relation.  It is therefore not consistent to prescribe an active stress \emph{a priori}, independent of the cell configuration.

Instead, we showed that a small change $\mathbf{g} \to \mathbf{g} + \delta \mathbf{g}$ of the tension metric behaves like an effective active stress acting on an (effective) elastic solid. 
Here, the ``rest state'' of the solid emerges from the current tension metric $\mathbf{g}$, and the effective active stress is given by $\tfrac12 \bm{\epsilon}^T\cdot \delta \mathbf{g} \cdot\bm{\epsilon}$.
In this sense, conventional active solids (active stress plus elastic ``background'') are recovered in the small strain approximation.
This again highlights the key prediction of our model that tissue deformation is caused by \emph{changes} in the active forces, not the active forces themselves. This has testable experimental consequences (as noted above and in the Introduction). In summary, the theory represents an approach to modeling morphogenesis distinct from existing ``active fluid'' models and recovers conventional ``active solid'' models in the small strain limit.

\paragraph*{Modeling tissue-scale flows with tension geometry.}

We hope that our framework will facilitate future studies of large-scale controlled tissue flow (e.g., primitive streak formation in the avian embryo~\cite{Caldarelli.etal2024} or convergent extension in \textit{Drosophila}~\cite{Streichan.etal2018}).
The continuum theory makes predictions in terms of fields that can be directly compared with experiments (e.g., tissue flow, which can be measured by particle image velocimetry, or motor-molecule anisotropy tensors~\cite{Streichan.etal2018}).

On the continuum level, a system like the primitive streak is described by a small number of linear PDEs, which are easy to solve numerically; cell-scale models must simulate $\sim 10^4 \mathrm{-} 10^5$ individual cells using custom software.
Further, the behavior of the continuum PDEs is likely reducible to a small number of modes. This may provide insight into how large-scale activity gradients and gene expression patterns control morphogenesis. For example, linear stability analysis can identify the drivers of large-scale tissue flow, which is very challenging in a cell-scale model.

So far, we have left the tension control tensor $\mathbf m$ unspecified. Practical applications will require a concrete \emph{model} for the tension dynamics based on hypothesized biological mechanisms (e.g., genetically patterned anisotropy~\cite{Pare.etal2014}, mechanical feedback loops~\cite{Fernandez-Gonzalez.etal2009}, or tension homeostasis). 
This takes the form of a dynamical equation for the tension-control tensor $\mathbf{m}$ as a function of variables already included in the theory (e.g., cell shape and tension anisotropy via the metrics $\mathbf{g}, \;\mathbf{a}$), or additional fields (e.g., morphogens).
Alternatively, the tension dynamics can be \emph{measured} using mechanical inference~\cite{Noll.etal2020}, and the tissue flow predicted from the ``empirical'' $\mathbf{m}$ can be compared to observations.

As a concrete example, let us consider positive tension feedback, implemented in the tension net framework in Refs.~\cite{Claussen.etal2024,Claussen.Brauns2025}.
In this model, tension triggers additional motor molecule recruitment with a rate $k_\mathrm{fb}$.
Tension anisotropy self-amplifies while the overall tension level remains fixed.
In terms of the tension control tensor, 
\begin{align}
    m_{zz} = 4 k_\mathrm{fb}\mu_{\zeta|z}, \quad m_{z\bar z} =0
\end{align}
On the other hand, T1s relax tension anisotropy by eliminating high-tension edges, as we saw in Sec.~\ref{sec:T1_relax}. Together,
\begin{align}
     \frac{d}{dt}\mu_{\zeta|z}=(k_\mathrm{fb}-\gamma_\mathrm{T1})\,\mu_{\zeta|z}
\end{align}
Before T1s set in, tension anisotropy grows and drives the QC tissue flow $\bar\partial_z\dot z= k_\mathrm{fb}\mu_{\zeta|z} $, Eq.~\eqref{eq:beltrami_flow_Eulerian}.
The tissue elongates perpendicular to the tension anisotropy direction (perpendicular to the high-tension interfaces), at a rate set by the feedback rather than by the absolute tension level.
T1s eventually decrease the anisotropy. A more involved mean-field calculation~\cite{Claussen.Brauns2025} shows that $\gamma_\mathrm{T1}>k_\mathrm{fb}$ at late times; hence, tension anisotropy decays and the flow is self-limiting.
Sustained flow requires a persistent oriented cue such as planar cell polarity, which may be represented by an additional field that contributes to the tension control tensor $m_{zz}$.

\paragraph*{Pressure effects and mechanical feedback.}

Cells sense not just tensions but also pressure~\cite{Coste.etal2010}. Interestingly, many tissues can avoid large pressure differentials and the potential instabilities they entail, like cell extrusion or rupture. On the cell level, pressures can be locally controlled via (negative) feedback loops\footnote{
The pressure feedback discussed here operates quasi-statically and requires a non-zero bulk modulus to build up pressure gradients. By contrast, in a compressible tissue, tension incompatibility causes secular elongation or contraction of cell-cell interfaces, requiring a different stabilizing mechanism (see Ref.~\cite{Noll.etal2017}).} coupling tensions and intracellular pressure.
We showed that pressure differentials are sourced by Gaussian curvature of the tension metric. Therefore, a minimal mathematical model for pressure-equilibrating feedback is $\partial_t \mathbf{g} = \beta \Delta\log (p/p_0) \,\mathbf{g} = -\beta K_g \,\mathbf{g} $, where $\beta$ is a feedback coefficient and we used Eq.~\eqref{eq:Gaussian_curvature}.  Cells can hence implement a ``flattening'' Ricci flow~\cite{Tao2008} by modulating their overall contractile activity in response to intracellular pressure. In contrast to, e.g., Frank elasticity in liquid crystals, Ricci flow only acts on the ``curvature part'' of $\mathbf{g}$ -- the tension manifold can freely remodel in-plane to drive morphogenetic flows.

On the tissue level, pressure can be controlled via feedback through cell proliferation and growth \cite{Shraiman2005,Gilbert.Barresi2016,Irvine.Shraiman2017},
which in our formulation acts on the target cell density $n(\xivec)$; Eq.~\eqref{eq:pressure-density-consistency}. 
Analyzing such ``adaptive'' feedback loops, and the combination of robustness and plasticity they confer, is an interesting direction for future research.

More generally, our findings have implications for how cells can sense and react to mechanical cues. We showed that the macroscopic stress is independent of the local tension configuration and, therefore, ``invisible'' at the cell level. Instead, the cell-level ``observables'' are the tension anisotropy $\mu_{\zeta | z}$ and the intracellular pressure $p$. 
Cell shape anisotropy could be sensed by intracellular structures such as the nucleus, intermediate filaments, or microtubules.

\subsubsection{Link to microscopic models}

Compared to microscopic, cell-scale models, as analyzed in the companion paper~\cite{Claussen.etal2026a}, the \textit{continuum} theory removes irrelevant, cell-scale details. It identifies the key ``hydrodynamic'' variables and how they determine large-scale tissue shape, stress, and flow.
Concretely, we saw that the tension metric defines an effective reference shape of the tissue, demonstrated an emergent elastic response and computed the effective moduli, identified the sources of macroscopic stress, and showed how local tension changes lead to tissue-scale (Beltrami) flow.
All of these insights rely on the continuum picture and cannot be read off directly from the discrete model.
Indeed, the central aim of the companion paper is to connect the discrete model to the present continuum theory, which can be understood and not just numerically simulated.

The companion paper~\cite{Claussen.etal2026a} confirms the results of the continuum theory by explicitly coarse-graining the ATN model, a specific example of a cell-scale ``vertex'' model. In particular, we recover the stress-metric relation (in the form Eq.~\eqref{eq:stress_w_cartesian}) by computing the coarse-grained stress via the Batchelor formula~\cite{Batchelor.Green1972}, and the mode decomposition of Sec.~\ref{sec:iso_conformal_decomp} from the cell-level force balance constraints.
On the cell scale, so-called \emph{discrete conformal maps}~\cite{Springborn.etal2008} take the place of the conformal maps and embeddings that occupy a central place in the continuum theory.
The Liouville equation Eq.~\eqref{eq:Liouville_pressure} becomes a (generalized) von Neumann law for the cell pressures.
Overall, the ``top-down'' analysis presented here is independently confirmed by ``bottom-up'' coarse-graining.

On a cell level, topological information on cell neighborhood relations can be ``geometrized'' via the so-called ``circle packing'' of the cell adjacency graph, whose existence and uniqueness are guaranteed by the Koebe--Andreev--Thurston theorem~\cite{Stephenson2005}.
The circle packing is a ``discrete'' conformal embedding of the equilateral triangulation defined by the adjacency graph.
These structures form the counterpart of the continuum adjacency metric and its isothermal coordinates.
In the discrete setting, the condition for T1 transitions can be calculated explicitly,
justifying the phenomenological T1-rate Eq.~\eqref{eq:T1_rate}.
Microstructural order can strongly influence T1 transitions, in which case additional order parameters become relevant \cite{Claussen.etal2024}. A continuum theory for these order parameters and their relation to topological defects in the cell packing is an interesting direction for future research.

Beyond the biological setting, the ATN model applies to
2d fluid or ferrofluid foams~\cite{Elias.etal1999, Weaire.etal2005}, the limiting case where all $\tau_{ij}=\tau=\mathrm{const}$. In such systems, dynamics could be introduced, for instance, through chemical reactions within foam bubbles, or by active control over bubble area~\cite{Kroo.etal2023}.

\paragraph*{Link to stress-first models of granular materials.}

In addition to foams, the ATN model is closely connected to another class of amorphous solids.
In \emph{granular materials}, one typically studies the statistical ensemble of compatible (force-balanced) force configurations for a given physical packing of particles~\cite{Henkes.Chakraborty2009,DeGiuli2018}.
This is the dual of the problem considered in the present study. 
Both frameworks build on microscopic force balance, but differ in how internal stress and observed physical configuration relate to one another. In granular media, there are many admissible states of self-stress for the same physical packing (contact network), and one is interested in calculating the probability of encountering a particular stress state.
In contrast, in a tension network, the mechanically balanced state is \textit{uniquely determined} by tensions and the constitutive equation for pressure\footnote{
If the pressure is instead constant, a multiplicity of balanced polygonal tilings (parametrized by the isogonal mode) exist for the same tension configuration, mirroring the degeneracy of stress states in granular materials.}. Changing tensions changes the physical configuration (cell positions and shapes).

As explained in the companion paper~\cite{Claussen.etal2026a}, there is an exact mathematical duality between tension nets and (frictionless) granular materials (as well as truss networks).
The duality exchanges the roles of ``force space'' and ``physical space''; the force network (tension triangulation) of an ATN gets mapped to the grain contact pattern of a granular material.

In future work, this duality might provide a starting point for continuum theories for granular matter. Indeed, previous phenomenological approaches postulate that granular materials can also be described by ``emergent elasticity''~\cite{Nampoothiri.etal2022}.

\subsubsection{Outlook}

While we have introduced our theory in the context of epithelial tissue mechanics, we believe it applies more broadly.
For example, protrusive mesenchymal cells have a more complex and fluctuating local architecture than confluent epithelia~\cite{Harris.etal1980,Oster.etal1983}. Nevertheless, 
their shape is also controlled by balancing active contractile stresses against bulk elasticity.
A tension metric could serve as a coarse-grained representation of dynamic local force generation more generally. The stress-metric relation might need to be modified to take into account differences in microscopic architecture (e.g., orientations of force dipoles) and how they change under material deformations. In the case considered here, the defining feature is that the microscopic forces are line tensions acting along interfaces orthogonal to the displacement vectors between adjacent cells. Phenomenologically, other stress-metric relations might be obtained based on symmetry arguments and dimensional analysis.

Lastly, one may ask if the tension metric approach can be generalized from two-dimensional tissue sheets to three-dimensional ``volumetric'' tissues. 
In 3d, the Riemannian geometry of the tension and adjacency metrics would become considerably richer~\cite{Lee2018}. For example, general 3d Riemannian metrics cannot be conformally mapped to flat Euclidean space, a key step in our theory.

To conclude, we believe that our theory provides a general framework for the emergent elasticity and plasticity of active solids, matter whose microscopic mechanics is not governed by conventional constitutive laws but by independently controlled active forces.
It quantitatively explains how these local active forces generate large-scale shape.

\begin{acknowledgments}
We thank D.~Cislo, A.~Košmrlj, and H.~Weyer for valuable feedback on the manuscript. B.I.S.\ acknowledges support of the NSF Physics (PoLS) grant \#2210612.
N.H.C.\ is supported by a fellowship from the Princeton Center for Theoretical Science, and by the National Institute of General Medical Sciences (NIGMS) of the National Institutes of Health (NIH) under award \#R01GM082938. 
F.B.\ acknowledges funding from the Max Planck Society and the Gordon and Betty Moore Foundation postdoctoral fellowship (grant \#2919).
The content is solely the responsibility of the authors and does not necessarily represent the official views of the National Institutes of Health or the National Science Foundation.

\end{acknowledgments}

\bibliography{GBE.bib}

\clearpage

\appendix
\setcounter{figure}{0}
\renewcommand{\thefigure}{S\arabic{figure}}

\begin{table}[h!]
    \renewcommand{\arraystretch}{1.1}
    \centering
    \begin{tabular*}{\linewidth}{l@{\extracolsep{\fill}}l}
    \toprule
    Symbol \hspace{3em} & Description \\
    \midrule
    $i,j,\dots$ & Cell indices \\
    $\alpha, \beta \dots$ & Coordinate indices \\
    $\tau_{ij}$ & Interfacial tension \\
    $\ell_{ij}$ & Cell edge lengths \\
    $p, p_0$ & Intracellular and reference pressure \\
    $n, \; n_0$ & Cell density and reference cell density \\
    $\delta n = n/n_0-1$ & Cell density deviation \\
    $P(n)$ & Pressure equation of state \\
    $B$  & Cellular bulk modulus \\
    $\hat{n}_\alpha$ & Unit normal \\
    $\epsilon_{\alpha\beta}$ & Antisymmetric tensor \\
    $(\dev q)_{\alpha\beta}$ & Deviatoric part symmetric matrix $q$ \\
    $\sigma_{\alpha\beta}, \; \sigma_{\alpha\beta}^\mathrm{tot}$ & Tensile and total stress tensor \\
    $\xi_{\alpha}$ & Lagrangian material coordinates \\
    $r_\alpha(\xi_1, \xi_2)$ & Physical configuration \\
    $h_{\alpha\beta}$ & Induced metric \\
    $g_{\alpha\beta}$ & Tension metric \\
    $a_{\alpha\beta}$ & Adjacency metric \\
    $G_{\alpha\beta}$ & Physical surface metric \\
    $K$ & Gaussian curvature of a metric \\
    $\Omega, \; \partial\Omega$ & Domain and boundary \\
    $z=z_1+iz_2$ & Complexified coordinates \\
    $q_{zz}, q_{z\bar z}$ & Complexified matrix components \\
    $\lambda_q$ & Conformal factor of symmetric matrix $q$ \\
    $\mu_q$ & Beltrami coefficient of symmetric matrix $q$ \\
    $\mu_{z|z'}$ & Beltrami coefficient of map $z\mapsto z'$ \\
    $\phi_{z|z'}$ & Phase of map $z\mapsto z'$ \\
    $\zeta$ & $\mathbf{a}$-isothermal coordinates \\
    $z$ & $\mathbf{g}$-isothermal reference state \\
    $w$ & Mechanically balanced state \\ 
    $u=w-z$ & Displacement field \\
    $Z$ & $\mathbf{G}$-isothermal physical surface coordinates  \\
    $\varepsilon_{\alpha\beta}$ & Strain tensor \\
    $\theta$ & Isogonal potential \\
    $\nu$ & Conformal displacement \\
    $t, \dot{\,}, dt$ & Time, time derivative and time increment \\
    $m_{\alpha\beta}$ & Tension control tensor \\
    $\eta$ & Cell growth factor \\
    $R_{\alpha\beta}$ & Oriented T1 rate \\
    $R_{\alpha\beta}^\mathrm{act}, R_{\alpha\beta}^\mathrm{pass}$ & Active and passive T1 rate \\
    $\gamma_{\mathrm{T1}}$ & T1 rate coefficient \\
    ATN & Active tension network \\
    EoS & Equation of state \\
    QC & quasiconformal \\
    BC & Boundary condition \\
    \bottomrule
    \end{tabular*}
    \caption{Overview of notation and abbreviations.}
    \label{tab:notation}
\end{table}

\section{Generalization to tissues on static curved surfaces}\label{app:curved}

In the main text, we considered tissues in a flat 2D plane.
The theory can be generalized to tissues confined to curved surfaces with fixed shape (defined, for example, by an eggshell, which balances the normal forces). The generalization to tissues that are free to deform out-of-plane is more complex and represents an interesting avenue for future work.

The physical surface $S$ on which the tissue lives is described by a metric $\mathbf{G}$, \emph{distinct} from the tension metric. In this Appendix, $\nabla_{\!\alpha}$ is the covariant derivative w.r.t. $\mathbf{G}$, and $\Delta=\nabla_{\!\alpha}\nabla^\alpha$. Physical configurations are now maps from the material manifold $\Omega$ to positions on the physical surface $S$.
We parametrize $S$ using a set of isothermal coordinates $Z$ for the physical metric $\mathbf{G}$ in which $G_{\alpha\beta} = \lambda_G^2(Z,\bar Z) \delta_{\alpha\beta}$. This coordinate system defines an orthogonal tangent basis $\lambda_G^{-1}  \partial_Z, \; \lambda_G^{-1}  \partial_{\bar{Z}}$ in which we express the local stress tensor (normalized so that $|\lambda_G^{-1}  \partial_Z|_G = \tfrac{1}{2}$). 
Since the physical surface is static, $\lambda_G$ will act like a fixed ``background field''.

\paragraph{Stress-metric relation.}

The derivation of the stress-metric relationship Eq.~\eqref{eq:stress-metric-relation} is fully local. It is therefore also valid in the tangent space of a curved surface $S$.
Eq.~\eqref{eq:stress_w} also generalizes and holds for the stress tensor components in the local basis (with the replacement $\lambda_g \rightarrow \lambda_g/\lambda_G$).

\paragraph{Effective reference state.}

Importantly, the physical metric $\mathbf{G}$ is defined on $S$, not the material manifold $\Omega$: tissue elements may be placed anywhere on $S$.
We use this freedom to find a conformal map from $(\Omega, \mathbf{g})$ to $(S, \mathbf{G})$.
Concretely, using the $\mathbf{g}$-isothermal coordinates $z(\xi,\bar\xi)$, we place a material element $\xi$ at physical position $Z = z(\xi, \bar\xi)$. 
Thus, $Z$ are a \emph{common} set of isothermal coordinates in which both physical and tension metrics are isotropic:
\begin{align}
    \mathbf{g} = \lambda_g(Z,\bar{Z} )^2 \,\mathbb{I}, \quad \mathbf{G} \ = \lambda_G(Z,\bar{Z} )^2 \,\mathbb{I}
\end{align}

By the stress-metric relationship, the tensile stress tensor in this configuration is isotropic, $\bm{\sigma} = p_0 (\lambda_g/\lambda_G) \mathbb{I}$. It can be balanced by a pressure $p=p_0 (\lambda_g/\lambda_G)$. Since $(\lambda_g/\lambda_G)$ is the conformal factor between $\mathbf{g}, \; \mathbf{G}$, the pressure is subject to the Liouville equation
\begin{align}
    \Delta \log (p/p_0) = K_G - (p/p_0)^2 K_g
\end{align}
where $K_G$ is the physical curvature, generalizing Eq.~\eqref{eq:Gaussian_curvature}.
Thus, mechanics is determined by the \emph{relative} curvature of the tension metric and physical surface. Overall, a stress-free reference state exists, just like in the flat case.

\paragraph{Adiabatic tissue dynamics.}

Since the effective reference state is defined by $Z=z(\xi,\bar\xi)$, the Beltrami flow due to adiabatic tension dynamics Eq.~\eqref{eq:beltrami_flow_Eulerian} is \emph{unaffected} by the physical curvature. Only the boundary conditions for the conformal factor change: in the curved case, they read $(\lambda_g/\lambda_G)\big|_{\partial\Omega} = 1$. As a result, the BC for the Beltrami flow becomes
$\Re\left[\lambda_G^{-2}\partial_Z(\lambda_G^2 \dot z) \right]\big|_{\partial\Omega}=\tfrac14 m_{z\bar z}\big|_{\partial\Omega}$.
This is simply the covariant version (using the Riemannian divergence) of the flat BC.

\paragraph{Force balance and elasticity.}

On a curved surface, in-plane force balance must be formulated using covariant derivatives, $\nabla_{\alpha}\sigma^\mathrm{tot}_{\alpha\beta}=0$~\cite{Salbreux.Julicher2017}. Effectively, the gradient of isotropic stress (pressure) must be replaced by the Riemannian gradient, and the divergence of anisotropic stress by the Riemannian divergence (with respect to the \emph{physical} metric $\mathbf{G}$).
The expression for the net in-plane force simplifies in isothermal coordinates:
\begin{align}
    \lambda_G^{-1}\left(
    \lambda_G^{-2}\partial_Z\big(\lambda_G^{2}\sigma_{ZZ}\big)+\bar\partial_Z\sigma_{Z\bar Z}-2\bar\partial_Z p
    \right) \stackrel{!}{=} 0 
\end{align}
Thus, physical curvature only enters via the coupling $2\lambda_G^{-1}(\partial_Z\lambda_G)\sigma_{ZZ}$ to the shear stress. Shear-free states (the isothermal reference and conformal deformations) are unaffected.
The linearized stress-strain relationship Eq.~\eqref{eq:effective_hooke} continues to hold, but the strain tensor must be computed using covariant derivatives, $\varepsilon_{\alpha\beta} \rightarrow \tfrac{1}{2}\left(\nabla_\alpha u_\beta +\nabla_\beta u_\alpha\right), \mathbb{I} \rightarrow \mathbf{G}$. 

On a curved space, covariant derivatives do not commute: $[\nabla^\alpha, \nabla^\beta\nabla_\beta]\theta = -K_G \nabla^\alpha\theta$. This modifies the equations for the effective Airy function
Eq.~\eqref{eq:PN_condition} and Eq.~\eqref{eq:Airy_linear}. Since describing curved surfaces is not our primary aim, we defer a detailed discussion of these effects to future work. In brief, $\delta n -B^{-1}\log(\lambda_g/\lambda_G)$ acts as a source term for residual stress, generalizing the flat case. 

\paragraph{Example.}
As an example, consider a tissue constrained to lie on a hemisphere of radius $1$. The physical isothermal coordinates $Z$ are defined by stereographic projection with $|Z|<1$ and $\lambda_G = 2/(1+|Z|^2)$.
By contrast, let us assume that the tension metric is the metric of a flat disk, $K_g=0$. Thus, the reference-state pressure is $p = p_0(\lambda_g/\lambda_G) = (1+|Z|^2)/2$. The tension per physical length, and with it the pressure, drops to $p_0/2$ at the pole.
The stress-free state is realized if the cell density matches the ``tension density'', so that the EoS $P(n)$ can supply the reference pressure. This requires lower cell densities at the pole. Otherwise, residual stresses will be generated.

\section{Examples for the static problem}{\label{app:examples}}

We provide two concrete examples to illustrate the continuum equations derived in the main text. Since the static problem is mathematically equivalent to linear elasticity, these examples serve to clarify the definitions of the tension- and adjacency metrics and the logic of the mapping from the tension manifold to the physical state.

\subsection{Statics: flat tension metric}{\label{app:example_flat}}

\begin{figure}
    \centering
    \includegraphics{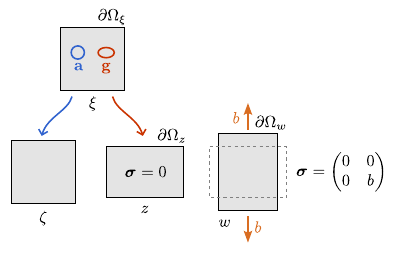}
    \caption{Elementary example for tension- and adjacency metrics illustrated by ellipses in the Lagrangian domain $\Omega_\xi$.}
    \label{fig:rect-example}
\end{figure}

As a first example, we revisit the rectangular foam from the Introduction, now in the continuum, as illustrated in Fig.~\ref{fig:rect-example}.
Let the adjacency metric be the Euclidean metric, corresponding to uniformly distributed isotropic cells. The $\mathbf{a}$-isothermal coordinates $\zeta$ are therefore simply Cartesian coordinates in the unit square $0\leq \zeta_1,\zeta_2 \leq 1$. In this square shape, the cell density is uniform, and cell shapes are isotropic. Next, we need to specify the active tension configuration, i.e., the tension metric $g$. In this example, we consider an anisotropic metric, $g_{\zeta\zeta} = 2s, \quad g_{\zeta\bar\zeta} = 2$ where the real parameter $0 < s \ll  1$ determines the strength of anisotropy. In Cartesian notation, $\mathbf{g}(\zeta_1,\zeta_2)= \mathrm{diag}[1+s, 1-s]$ corresponding to $\tau_1^2 = 1-s$, $\tau_2^2 = 1+s$ in the rectangular foam discussed in the Introduction.
Microscopically, this tension metric means that vertical cell-cell interfaces are under higher tension than horizontal ones, since the tension metric encodes the tension on interfaces transverse to a line element $d\rvec$ (cf.\ Fig.~\ref{fig:virtual_cut}).
The Beltrami coefficient reads $\mu_g = s/(1+\sqrt{(1+s)(1-s)}) \approx s/2$, where the approximation is for $s \ll 1$. To remove this anisotropy, the $\mathbf{g}$-isothermal coordinates need to create a corresponding shear, $z(\zeta,\bar\zeta) = \zeta + \mu_g \bar\zeta$. This fulfills the Beltrami Eq.~\eqref{eq:beltrami_g}. In Cartesian notation, $\mathbf{z} = \mathrm{diag}[1+\mu_g, 1-\mu_g]\cdot \bm{\zeta}$.

Since $\det \mathbf{g} \approx 1$, tension and cell densities are (approximately) equal in this example. Further, since $\mathbf{g}$ is constant, it has no curvature, $K_g=0$: the $z$-configuration has constant cell density and constant pressure $p_0$. It is therefore mechanically balanced in the absence of external forces. In other words, $u_\alpha =0$ solves the linear elasticity force-balance equation Eq.~\eqref{eq:navier}.
This example illustrates how tension anisotropy controls tissue shape.

Let us now consider applying a constant boundary traction force $b_1=0, b_2=\pm b$ to the upper ($+$) and lower ($-$) edges of the rectangular tissue patch (the left and right sides remain free). This results in a potential-conformal displacement $u_\alpha = \nu_\alpha + \partial_\alpha \theta$ from the tension reference $z$ to the physical configuration $w=z+u(z,\bar z)$. By Eq.~\eqref{eq:Airy_linear}, the potential $\theta$ must be biharmonic, $\Delta^2 \theta = 0$.
The traction forces $\sigma\cdot \mathbf{n}_{\pm} = [0, \pm b]$ at the top and bottom edges require $\sigma_{22}|_{\pm} = -p_0 (\partial_1^2 \theta)|_{\pm} = b$. We can hence take $\theta = -b \Re[z]^2/(2 p_0)$, which implies a uniform stress $\bm{\sigma} = \mathrm{diag}\,(0, b)$ in the bulk and, thus, also fulfills the free BCs on the left and right edges.
This result is expected by simple physical reasoning.
To ensure that the displacement $u_\alpha$ is incompressible, we add a conformal contribution $\nu(z) = b z/(2p_0)$ (in this case, simply a uniform scaling).

\subsection{Statics: non-flat tension metric}{\label{app:hemisphere}}

\begin{figure}
    \centering
    \includegraphics{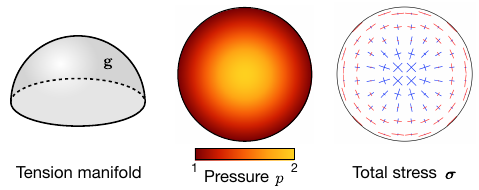}
    \caption{A half-sphere tension manifold induces a non-constant pressure field and a non-vanishing total stress even for traction-free boundary conditions. Note that the total stress is isotropic in the center and purely tangential along the boundary, as is required by the traction-free boundary conditions.}
    \label{fig:example-stress}
\end{figure}

Next, we study an example where the tension metric has curvature. 
We consider a circular tissue patch with material coordinates $\xi_1,\xi_2$ covering the unit disk.
We now define the tension metric by:
\begin{align}
    \renewcommand{\arraystretch}{1.4}
    (g_{\alpha\beta}) = \begin{pmatrix}
        1+\xi_1^2/h^2 & \xi_1\xi_2/h^2 \\
        \xi_1\xi_2/h^2 & 1+\xi_2^2/h^2
    \end{pmatrix}, \; h(\xi,\bar\xi)^2 = 1-\xi\bar\xi
\end{align}
We use round brackets to highlight that this matrix uses Cartesian, not complexified, coordinates.
It is the metric of a unit hemisphere, a height $h$ above the unit disk in Monge parametrization (Fig.~\ref{fig:example-stress}). 
Since $\det\mathbf{g} = 1/h(\xi,\bar\xi)^2$, the ``tension density'' is higher at the margin than at the center.
(We note that this pedagogical example involves large deformations, so the results of the linearized formalism used below may be inaccurate.)

We now need to find the transformation $z(\xi,\bar\xi)$ that makes this tension metric isotropic. Because of rotation symmetry, we expect a radial dilation. The $g$-isothermal coordinates are given by the stereographic projection~\cite{Ahlfors1979} of the hemisphere, which is a conformal map:
\begin{equation}
    z(\xi,\bar\xi) = \frac{\xi}{1+h(\xi,\bar\xi)}
\end{equation}
The conformal factor reads $\lambda_g(z,\bar z) = 2/(1+z\bar z)$, which fulfills $\lambda_g = 1$ on the boundary. By Eq.~\eqref{eq:Gaussian_curvature}, the curvature is constant, $K_g = 1$, as expected for a unit sphere.

Due to curvature, area distortion (and thus cell density gradients) in the tension reference state $z$ are unavoidable. For an incompressible tissue, these must be compensated by an isogonal-conformal displacement $w -z = u = f(z) + 2\bar\partial_z\theta$; Eq.~\eqref{eq:navier}.

We assume that the cell density is uniform w.r.t. the tension metric, i.e., $n(z,\bar z) =n_0\lambda_g^2(z,\bar z)/2$. Incompressibility then means $(\det \partial_{\mathbf{z}} w)^{-1} \sqrt{\det g(z,\bar z)} =2$ where $(\partial_{\mathbf{z}} w)_{\alpha\beta} = \partial w_\beta / \partial z_\alpha $ is the Jacobian. Expanding $\det \partial_{\mathbf{z}}w = \det[\delta_{\alpha\beta} + \partial_\alpha u_\beta]$ with $u_\alpha = \nu_\alpha +\partial_\alpha \theta$ to linear order:
\begin{equation}
    \Delta\theta +\partial_\alpha \nu_\alpha = \frac{2}{(1+z\bar z)^2} -1
    \label{eq:harmonic_hemisphere}
\end{equation}
We now obtain a rotation-symmetric solution to Eq.~\eqref{eq:harmonic_hemisphere} as a function of $r=\sqrt{z\bar z}$.
First, the only rotation-symmetric conformal mode is a uniform scaling $f(z)=Cz$. 
Further, we fix an irrelevant global constant by setting $\theta(0) = 0$.
The equation for $\theta(r)$ then reads
\begin{equation}
    \label{eq:harmonic_hemisphere_radial}
    \Delta \theta(r) + 1 = \frac{2}{(1+r^2)^2}
\end{equation}
with BCs $\theta(0) = 0$, $\partial_r \theta(r=0) = 0$, and $\partial_r \theta(r=1) = 0$. The last condition follows from $\sigma\cdot\mathbf{n}=0$ (vanishing traction forces at the boundary), since $\sigma_{rr}=-p_0(\partial_r\theta)/r$. Eq.~\eqref{eq:harmonic_hemisphere_radial} is solved by 
\begin{align}
    \theta(r) = \tfrac{1}{4}[-r^2 + 2 \log(1+r^2)]
\end{align}
We conclude that $C = 0$. No conformal displacement is needed, because the total area of the reference state already matches the cell number.

Within the disk, $\sigma_{\alpha\beta}^\mathrm{tot} = -p_0 \epsilon_{\alpha\gamma}\epsilon_{\beta\delta}\partial_\gamma\partial_\delta \theta\neq 0$; Eq.~\eqref{eq:Airy_stress_continuum}. 
These internal residual stresses, shown in Fig.~\ref{fig:example-stress}, result from curvature of the tension manifold.
Importantly, such stresses will also be present for finite cell compressibility. In the limit of fully compressible cells, no force-balanced configuration exists because sustaining the pressure field implied by the tension manifold curvature requires a finite cell compressibility. 

\section{Change of curvature radius by a conformal map}{\label{app:curvature_conformal}}
Here, we compute the change of a line's curvature radius $R$ by a conformal map $z\mapsto f(z)$~\cite{Ahlfors1979}. Consider a line $z(t)$ with normal ${\mathbf{n}}$. The rotation of the local normal is given by the local rotation angle $\omega = \arg\partial_z f$. Moving a step $dt$ along the transformed curve $f(z(t))$, the tangent $i\mathbf{n}$ rotates by an angle $d\phi = \omega(z(t+dt)) - \omega(z(t)) = (i {\mathbf{n}}) \cdot 2 \bar \partial_z \omega dt$. Due to the scaling factor $|\partial_z f|$, the arc length is $|\partial_z f| dt$. Altogether, the curvature reads:
\begin{align}
    \frac{1}{R} &= \frac{d\phi}{d(|\partial_z f| t)} = \frac{i{\mathbf{n}}}{|\partial_z f|} \cdot 2\bar\partial_z \arg\partial_z f = \frac{\mathbf{n}}{|\partial_z f|^2} \cdot 2\bar \partial_z |\partial_z f|
\end{align}

\section{The isogonal-conformal decomposition}\label{app:isogonal_conformal_decomp}

Here, we show that any vector field $u_\alpha$ that fulfills the solvability condition $\Delta(\epsilon_{\alpha\beta}\partial_\beta u_\alpha) = 0$ (the vorticity is harmonic) can be decomposed as  $u_\alpha=\partial_\alpha \theta + \nu_\alpha$, where $\nu_\alpha$ is a conformal vector field. In contrast to the familiar Helmholtz decomposition into curl- and divergence-free parts, this decomposition is only possible if the solvability condition is fulfilled.
For example, a solenoidal vector field $u_\alpha=\epsilon_{\alpha\beta}\partial_\beta \psi$ must have $\Delta^2 \psi = 0$.

In complex notation, given $u$, we search for a real $\theta$ so that 
$u-2\bar\partial\theta$ is holomorphic, i.e., $\bar\partial u-2\bar\partial^2 \theta =0$. Since $\theta$ is real, applying $\partial^2$ leads to the solvability condition 
\begin{align}
    0 =\Im[\partial^2\bar\partial^2 \theta] = \Im[\partial^2\bar\partial u]= \tfrac18\Delta(\partial_1u_2-\partial_2u_1)
\end{align}
which is precisely the assumption of harmonic vorticity.
We can thus find a potential $\theta_0$ so that $\partial^2 \bar\partial u-2\partial^2\bar\partial^2 \theta_0=0$.
The residual satisfies $\partial^2 \left(\bar\partial u-2\bar\partial^2 \theta_0\right) = 0$ and may thus be written as $\bar\partial u-2\bar\partial^2 \theta_0 = a(\bar{z}) +z b(\bar z)$. 
The potential $\theta_1 = A(\bar z)+zB(\bar z)+ A(z) +\bar z B(z)$ with $A'' = a, \; B'' =b$ thus completes the desired solution, i.e., $\bar\partial u - 2\bar\partial^2 (\theta_0+\theta_1) = 0$.

\section{Boundary conditions}\label{app:boundary_conditions}

\paragraph{Conformal deformations cannot accommodate arbitrary boundary deformations.}
A flat conformal map $f$ is fully specified by its conformal factor $|\partial_z f|$, which obeys the Liouville equation $\Delta \log |\partial_z f|=0$. This Laplace problem admits prescribing either Dirichlet conditions $|\partial_z f||_{\partial\Omega}$, or Neumann conditions $\partial_\mathbf{n}|\partial_z f||_{\partial\Omega}$ (the boundary curvature), but not both simultaneously.
Equivalently, one can specify BCs for one of the two components $f_\alpha$ -- the second component is fixed by the Cauchy-Riemann equations $\partial_\alpha f_1 = \epsilon_{\alpha \beta} \partial_\beta f_2$.

For example, consider the unit disk in complex polar coordinates $z=re^{i\phi}$. Conformal maps $f$ are analytic, $f(z) =\sum_{n\geq 0} \tfrac{f^{(n)}(0)}{n!} z^n$. Importantly, the sum is restricted to $n\geq 0$: hence, the boundary value $f(e^{i\phi})$ has only non-negative Fourier coefficients. 
A boundary displacement with negative Fourier coefficients cannot be extended to a conformal deformation of the disk. For example, pure shear $\mathbf{u}|_{\partial\Omega} \propto e^{-i\phi} =x \hat{\mathbf{x}} - y \hat{\mathbf{y}}$ is incompatible with a conformal deformation.

\paragraph{Boundary conditions for the Beltrami equation.}
In the context of the Beltrami equation Eq.~\eqref{eq:beltrami_g}, the BCs fix the conformal contribution to $z(\xi, \bar\xi)$. Therefore, the Beltrami equation also admits a single BC only, e.g., the natural BC for $|\partial_\xi z|$, Eq.~\eqref{eq:natural_bcs}. Indeed, the example in App.~\ref{app:example_flat} makes it clear that it is impossible to make the tension metric isotropic while keeping the boundaries clamped.
The Beltrami equation must be contrasted with the elasticity problem for the physical configuration $w$, which admits two boundary conditions.

\section{Kinematics of quasiconformal flow}
\label{app:beltrami-kinematics}

\begin{figure}[ht]
    \centering
    \includegraphics{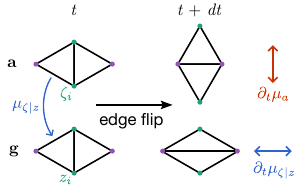}
    \caption{Illustration of the geometry of Eq.~\eqref{eq:Beltrami-convective-app}.
    The adjacency metric represents a triangulation where each edge has length 1, and therefore undergoes a shear deformation if adjacency is changed by an edge flip (T1 transition). The tension metric remains fixed under a T1. As a result, the tension triangle shape acquires anisotropy.
    }
    \label{fig:Beltrami-flip}
\end{figure}

\subsection{Convective derivative of $\mu_{\zeta|z}$}
\label{app:beltrami-kinematics_1}

Consider the QC flow $\zeta' = \zeta + {\dot{\zeta}} dt$ under dynamics of the adjacency metric; $\bar{\partial}_\zeta {\dot{\zeta}} = \partial_t \mu_a$.
We calculate how $\mu_{\zeta|z}$ changes under this flow while holding the tension metric $\mathbf{g}$ fixed.
To find the new Beltrami coefficient $\mu_{\zeta'|z}$, we need to compose the maps $\zeta' \mapsto \zeta$ and $\zeta \mapsto z$. 
Drawing the map compositions as a diagram
\begin{equation}
\begin{tikzpicture}[scale=0.75,baseline]
    \node (zeta) at (0,0) {$\zeta$};
    \node (zeta') at (4,0) {$\zeta'$};
    \node (z) at (6,0) {$z$};
    \draw [red, ->, bend left, line width=0.75pt] (zeta') edge node[anchor=north, inner sep=2pt] {$\mu_{\zeta'|\zeta}$} (zeta);
    \draw [red, ->, bend angle=60, bend left, line width=0.75pt] (zeta) edge node[anchor=north, inner sep=3pt] {$\mu_{\zeta|z}$} (z);
    \draw [->, line width=0.75pt] (zeta) edge node[anchor=south, inner sep=2pt] {$\mu_{\zeta|\zeta'}=\partial_t \mu_a dt$} (zeta');
    \draw [->, line width=0.75pt] (zeta') edge node[anchor=north, inner sep=2pt] {$\mu_{\zeta'|z}$} (z);
\end{tikzpicture}
\end{equation}
QC flow induces $\mu_{\zeta|\zeta'} = \bar{\partial}_\zeta {\dot{\zeta}} dt / (1 + \partial_\zeta {\dot{\zeta}} dt) \approx \bar{\partial}_\zeta {\dot{\zeta}} dt = \partial_t \mu_a dt$.
To find the Beltrami coefficient of the inverse of a map, one applies the chain rule:
\begin{equation} \label{eq:Beltrami-inverse}
    \mu_{\zeta'|\zeta} = - \frac{\partial_\zeta \zeta'}{\overline{\partial_\zeta \zeta'}} \mu_{\zeta|\zeta'}.
\end{equation}
Now composing with the map $\zeta \mapsto z$ gives
\begin{align}
    \mu_{\zeta'|z}&(\zeta', \bar\zeta')
    = \frac{\mu_{\zeta'|\zeta} + \mu_{\zeta|z} \, e^{-i2\phi_{\zeta'|\zeta}}}{1 + \bar{\mu}_{\zeta'|\zeta} \, \mu_{\zeta|z} \, e^{-i2\phi_{\zeta'|\zeta}}}   \\
    &\approx \frac{-\bar{\partial}_\zeta {\dot{\zeta}} dt + \mu_{\zeta|z} (1 + i 2 \mathrm{Im}[\partial_\zeta {\dot{\zeta}}] dt)}{1 - \mu_{\zeta|z} \partial_\zeta \overline{\dot{\zeta}} dt} \nonumber \\
    &\approx \mu_{\zeta|z} - \left( \partial_t \mu_a - 2i \mu_{\zeta|z} \mathrm{Im}[\partial_\zeta {\dot{\zeta}}] - \mu_{\zeta|z}^2 \partial_t \bar{\mu}_a \right) dt, \nonumber
\end{align}
where the phase factor is $e^{-i2\phi_{\zeta'|\zeta}} = (1 - \overline{\partial_\zeta {\dot{\zeta}}} dt)/(1 - \partial_\zeta {\dot{\zeta}} dt) = 1 + i 2 \mathrm{Im}[\partial_\zeta {\dot{\zeta}}] dt$.
The displacement change from $\zeta$ to $\zeta'$ leads to an additional advective term $\mathrm{Re}[\bar{\dot{\zeta}} \partial_\zeta]\mu_{\zeta |z} dt$. 
Taken together, one arrives at
\begin{align} \label{eq:Beltrami-convective-app}
    \frac{d\mu_{\zeta|z}}{dt} 
    &=\frac{\mu_{\zeta'|z}(\zeta', \bar\zeta') - \mu_{\zeta|z}(\zeta, \bar\zeta)}{dt} \\
    &=\partial_t \mu_g - \partial_t \mu_a 
    \nonumber \\ & \quad
    + i 2 \mu_{\zeta|z} \mathrm{Im}[\partial_\zeta {\dot{\zeta}}] + \mu_{\zeta|z}^2 \partial_t \bar{\mu}_a
\end{align}
The final, nonlinear term can be neglected for weak anisotropy $|\mu_{\zeta|z}|\ll 1$.
In conventional Cartesian notation, the total derivative of the tension metric $\mathbf{g}$ can be written as 
\begin{align}\label{eq:advection}
    \frac{d\mathbf{g}(\zeta,\bar\zeta)}{dt} = \mathbf{g} \mathbf{m} - \left[({\dot{\zeta}} \cdot \nabla)\mathbf{g} +  \nabla{\dot{\zeta}} \cdot \mathbf{g} + \mathbf{g}\cdot (\nabla{\dot{\zeta}})^T \right]
\end{align}
Figure~\ref{fig:Beltrami-flip} provides a geometric explanation of Eq.~\eqref{eq:Beltrami-convective-app}.
The term $\partial_t \mu_g$ arises from the intrinsic dynamics of $z$, in addition to the coordinate transformation $\zeta\mapsto\zeta'$.

\subsection{Variation of $|\mu_{\zeta|w}|^2$}{\label{app:beltrami-kinematics_2}}

In the main text, we argue that T1s relax the ``total'' Beltrami coefficient $\mu_{\zeta|w}$ of the map from ${\bf a}$-isothermal coordinates $\zeta$ to the physical state $w$.
We therefore need to calculate the variation of $\mu_{\zeta|w}$ under a small variation of the adjacency metric ${\bf a}(\bm{\zeta}) \mapsto {\bf a}'(\bm{\zeta})= {\bf a}(\bm{\zeta})+\delta \mathbf{a}(\bm{\zeta})$.
First, note that $\mu_{\zeta|w}$ is independent of $\mathbf{g}$ since the map $\zeta \mapsto w$ bypasses the intermediate (reference) configuration defined by $\mathbf{g}$ (at fixed $w$, cf. Fig.~\ref{fig:maps}). 
Now let $\zeta'$ be the $a'$-isothermal coordinates, which must satisfy the Beltrami equation $\bar \partial_\zeta \zeta' = \delta \mu_{a} \, \partial_\zeta \zeta'$.
Drawing the map compositions as a diagram
\begin{equation}\label{eq:a-variation-maps}
\begin{tikzpicture}[scale=0.75,baseline]
    \node (zeta) at (0,0) {$\zeta$};
    \node (zeta') at (4,0) {$\zeta'$};
    \node (w) at (6,0) {$w$};
    \draw [red, ->, bend left, line width=0.75pt] (zeta') edge node[anchor=north, inner sep=2pt] {$\mu_{\zeta'|\zeta}$} (zeta);
    \draw [red, ->, bend angle=60, bend left, line width=0.75pt] (zeta) edge node[anchor=north, inner sep=3pt] {$\mu_{\zeta|w}$} (w);
    \draw [->, line width=0.75pt] (zeta) edge node[anchor=south, inner sep=2pt] {$\mu_{\zeta|\zeta'}=\delta \mu_{a}$} (zeta');
    \draw [->, line width=0.75pt] (zeta') edge node[anchor=north, inner sep=2pt] {$\mu_{\zeta'|w}$} (w);
\end{tikzpicture}
\end{equation}
makes clear that the Beltrami coefficient of the new map $\zeta'\mapsto w$ is calculated by composing $\zeta' \mapsto \zeta \mapsto w$ [red path in Eq.~\eqref{eq:a-variation-maps}].
We first use that $\zeta \mapsto \zeta'$ is simply the QC flow induced by $\delta \mathbf{a}$ so $\mu_{\zeta|\zeta'} = \bar\partial_{\zeta} \zeta' / \partial_{\zeta} \zeta' = \delta \mu_a$.
The inverse is obtained using Eq.~\eqref{eq:Beltrami-inverse}, giving
\begin{equation}
    \mu_{\zeta'|\zeta} = -\frac{\overline{\partial_{\zeta'} \zeta}}{\partial_{\zeta'} \zeta} \mu_{\zeta|\zeta'} = - \frac{\overline{\partial_{\zeta'} \zeta}}{\partial_{\zeta'} \zeta} \delta \mu_a    
\end{equation}
Since $|\mu_{\zeta'|\zeta}| \ll 1$, the Beltrami coefficient of $\zeta' \mapsto \zeta \mapsto w$ can be calculated by linearizing the Beltrami composition formula Eq.~\eqref{eq:Beltrami_composition}:
\begin{align}
    \mu_{\zeta'|w} 
    &\approx \mu_{\zeta'|\zeta} + \frac{\overline{\partial_{\zeta'} \zeta}}{\partial_{\zeta'} \zeta} \mu_{\zeta|w}
    = \frac{\overline{\partial_{\zeta'} \zeta}}{\partial_{\zeta'} \zeta} \left( \mu_{\zeta|w} - \delta \mu_a \right)
\end{align}
And therefore, we conclude:
\begin{align}
    |\mu_{\zeta'|w}|^2 &= |\mu_{\zeta|w}|^2 - (\mu_{\zeta|w} \delta \bar{\mu}_a + \bar{\mu}_{\zeta|w} \delta\mu_a)   \\[0.6em]
    \Rightarrow 
    \frac{\delta |\mu_{\zeta|w}|^2}{\delta \bar{\mu}_a} &= -\mu_{\zeta|w}.
\end{align}

\end{document}